\documentclass[twocolumn,amsmath,amssymb,floatfix,prd,nofootinbib]{revtex4-1}
\usepackage{epsfig}
\usepackage{dcolumn}
\usepackage{bbold} %
\usepackage{wasysym} %
\usepackage{hyperref}
\usepackage{csquotes}
\usepackage{slashed}
\usepackage{hepnames}

\usepackage{hyperref}
\usepackage{tabularx}
\usepackage{siunitx} %

\DeclareMathOperator{\DiLog}{\text{Li}_2}
\DeclareMathOperator{\tQED}{\text{QED}}
\DeclareMathOperator{\tOK}{\text{OK}}

\DeclareMathOperator{\Md}{\mathcal M}

\DeclareRobustCommand{\PQ}{\HepGenParticle{Q}{}{}\xspace} %
\DeclareRobustCommand{\PaQ}{\HepGenAntiParticle{Q}{}{}\xspace} %

\begin{document}
\title{Next-to-Leading Order QCD Corrections to Inclusive Heavy-Flavor Production\newline
in Polarized Deep-Inelastic Scattering}
\author{Felix Hekhorn}
\email{felix.hekhorn@uni-tuebingen.de}
\affiliation{Institute for Theoretical Physics, University of T\"ubingen, Auf der Morgenstelle 
14, 72076 T\"ubingen, Germany}
\author{Marco Stratmann}
\email{marco.stratmann@uni-tuebingen.de}
\affiliation{Institute for Theoretical Physics, University of T\"ubingen, Auf der Morgenstelle 
14, 72076 T\"ubingen, Germany}

\begin{abstract}
We provide a first calculation of the complete next-to-leading order QCD corrections for heavy flavor
contributions to the inclusive structure function $g_1$ in longitudinally polarized deep-inelastic
scattering. The results are derived with largely analytical methods and retain the full dependence
on the heavy quark's mass. We discuss all relevant technical details of the calculation and present numerical
results for the heavy quark scaling functions. 
We perform important crosschecks to verify our results in the known limit of photoproduction and
for the unpolarized electroproduction of heavy quarks.
We also compare our calculations to the available, partial results in the polarized case,
in particular, in the limit of asymptotically large photon virtualities, and analyze the behavior of
the scaling functions near threshold.
First steps towards phenomenological applications are taken by providing some estimates 
for inclusive charm production in polarized deep-inelastic scattering at a 
future electron-ion collider and studying their sensitivity to the polarized gluon distribution. 
The residual dependence of heavy quark electroproduction on unphysical factorization and renormalization scales 
and on the heavy quark mass is investigated.
\end{abstract}

\maketitle

\section{Introduction and Motivation}
Heavy quarks (HQ) are an important and versatile laboratory for 
probing different aspects of Quantum Chromodynamics (QCD) 
\cite{Baines:2006uw} ranging from the heavy flavor content of nucleons and the hadronization of
heavy quarks into heavy mesons or baryons to particular details of the dynamics of
QCD hard scattering processes.
The heavy quark mass $m$ acts as a natural regulator in perturbative 
calculations of otherwise singular kinematic configurations
and renders, for example, the notion of a total heavy quark
production cross section meaningful.
On the other hand, keeping the full dependence on the heavy quark's mass
throughout higher order calculations within perturbative QCD (pQCD)
significantly complicates, for instance, 
analytical phase-space integrations. Unlike for massless QCD processes,
HQ cross sections start to depend on at least
two energy scales, the HQ's mass and some other kinematic quantity characterizing the
process such as the virtuality $Q$ of the exchanged virtual photon $\Pggx$ 
in deep-inelastic scattering (DIS) or the HQ's transverse momentum 
$p_T$ in hadron-hadron collisions.

Multi-scale problems demand extra care as they may require all-order 
resummations in certain regions of phase-space depending on the hierarchy 
of the relevant scales.  
In case of HQ production in DIS, which we consider in this paper,
this is intimately linked with the question of how to define and treat HQ parton
densities properly in the entire range of $Q^2$ \cite{Butterworth:2015oua,Accardi:2016ndt}
from the threshold, $Q^2\simeq m^2$, to the asymptotic regime, $Q^2\gg m^2$.
In general, one deals with the question of how to match a
theory with $n_{lf}$ light quark flavors and one HQ to a theory with $n_{lf}+1$ massless flavors
by resumming logarithms of the type $\alpha_s^l\,\ln^k(Q^2/m^2)$, $1\le k\le l$
in each order of perturbation theory. Thereby one introduces a HQ parton density at some matching scale $\mu$. 
In order to achieve a unified description for HQ production for all values of $Q^2$, 
different types of general-mass variable flavor number schemes (GM-VFNS) 
have been proposed \cite{Butterworth:2015oua,Accardi:2016ndt}
and adopted in the various global fits of unpolarized parton distribution functions (PDFs)
that are currently available. No such schemes have been considered and invoked so far
in case of helicity-dependent PDFs that can be accessed, for instance, in the DIS  
of longitudinally polarized lepton beams off longitudinally polarized nucleons.
The kinematics of existing data \cite{Aidala:2012mv} is such that 
even the charm contribution is negligible, albeit nonzero, 
and, hence, sets of helicity PDFs are usually extracted with either $n_{lf}=3$ light quarks only 
or within a naive zero-mass variable flavor number scheme (ZM-VFNS) that neglects all
mass effects for heavy flavors \cite{deFlorian:2008mr,Blumlein:2010rn}.
In addition, the required theoretical expressions for HQ production in helicity-dependent
DIS with the full dependence on the HQ mass are still lacking, a gap that will be closed in this paper.

In quantitative, phenomenological studies of the nucleon structure in terms of PDFs, 
data on HQ production draw their particular relevance from the pronounced
dominance of gluon-induced production processes already at the lowest order (LO) approximation in pQCD.
In case of DIS, only photon-gluon fusion (PGF) contributes at the Born approximation, which
makes it particularly sensitive to the gluon distribution. Therefore,
data on the charm contribution to the DIS structure function $F_2$ taken at
the DESY-HERA lepton-proton collider \cite{Abramowicz:1900rp}
are utilized in all global analyses of unpolarized PDFs \cite{Butterworth:2015oua,Accardi:2016ndt}. 
At small momentum fractions $x$ the charm contribution
to the DIS cross section amounts to about $25\%$ and, hence, must be treated
properly in phenomenological analyses of PDFs, i.e., at least 
the full next-to-leading order (NLO) QCD corrections are required.

In case of spin-dependent PDFs \cite{deFlorian:2008mr,Blumlein:2010rn}, 
the gluon helicity density $\Delta \Pg(x,Q^2)$ is still completely 
unconstrained at low momentum fractions $x$ due to the lack of data,
which prevents one from answering one of the most topical questions in Nuclear Physics, namely 
what is the net contribution of gluons to the spin of the proton, i.e.,
what is the value of its first moment $\int_0^1 \Delta \Pg(x,Q^2)\,dx$.
Recent data from polarized proton-proton collisions at BNL-RHIC \cite{Aschenauer:2015eha}
have revealed first evidence for a sizable contribution to the integral at medium-to-large
values of $x$ \cite{deFlorian:2014yva}, but nothing can be said about $\Delta \Pg(x,Q^2)$
for $x$ values smaller than about $0.01$. Among other measurements, data on
HQ production in polarized DIS at small $x$, i.e., to the relevant spin-dependent structure function
$g_1(x,Q^2)$, would prove to be very useful in addressing this question further.
The planned, high-luminosity Electron-Ion Collider (EIC) in the U.S.~\cite{Boer:2011fh}, whose physics case and
technical realization is currently under scrutiny, would uniquely offer, for the first time,
access to a broad kinematic regime of small-to-medium momentum fractions $x$ in a range of $Q^2$,
where heavy quark, in particular, charm, contributions to $g_1$ in polarized DIS 
could be sizable and, hence, experimentally accessible within meaningful uncertainties.

Expected helicity-dependent data from the EIC 
in the phenomenologically relevant small-$x$ regime, say, below $x\simeq 0.01$
would be available at rather modest values of $Q^2$, i.e., 
far from the asymptotic regime $Q^2\gg m^2$ for HQ production in DIS.
Hence, to describe HQ electroproduction in polarized DIS at an EIC reliably in pQCD 
the exact dependence on the HQ mass $m$ must retained. In addition, 
the full NLO QCD corrections have to be computed and taken 
into account in quantitative analyses. 
In general, NLO corrections to processes involving heavy quarks are
often known to be sizable and not uniform in the 
relevant kinematic variables $x$, $Q^2$, or $p_T$.
In any case, they are needed to accurately estimate HQ yields 
quantitatively and to turn them into useful constraints onto PDFs
by reducing, e.g., the theoretical uncertainties associated with unphysical 
renormalization and factorization scales.

In this paper we complete the suite of NLO calculations of heavy flavor production
with longitudinally polarized beams and targets. While complete NLO results for both
photo- \cite{Bojak:1998bd,Merebashvili:2000ya,Riedl:2012qc}
and hadroproduction \cite{Bojak:2001fx} are available in the literature for quite some time now, 
NLO corrections for the heavy contribution to the inclusive DIS structure function $g_1$ 
are still lacking apart from the known analytic expressions in the asymptotic limit $Q^2\gg m^2$ \cite{Buza:1996xr}
and analytic results for the genuine NLO Compton-like light quark-induced
subprocess \cite{Buza:1996xr,Blumlein:2016xcy}. 
We note that the LO expression for polarized photon-gluon fusion has been computed 
in \cite{Watson:1981ce,Vogelsang:1990ug} a long time ago.

To perform our NLO calculations that retain the full dependence on the HQ mass $m$,
we follow closely the semi-analytical methods used in the
computation of the corresponding unpolarized inclusive HQ electroproduction 
at NLO accuracy in Ref.~\cite{Laenen:1992zk}. Angular phase-space integrations for the partonic subprocesses
are performed largely analytically but, in general, two remaining integrations 
as well as the convolution with the helicity parton densities have to be done numerically.
Singularities in intermediate steps of the calculations are made manifest in dimensional regularization
as poles in $1/\epsilon^2$ and $1/\epsilon$, i.e., we choose to work in
$n=4+\epsilon$ dimensions and only take the limit $\epsilon\to 0$ in the end to arrive at
the final, finite expressions for the NLO HQ coefficient functions for polarized inclusive
electroproduction. The Dirac matrix $\gamma_5$ and the Levi-Civita tensor that appear throughout the calculations
are dealt with in the commonly used 't~Hooft-Veltman-Breitenlohner-Maison (HVBM) prescription \cite{tHooft:1972tcz}
which leads to some well-known subtleties in $n$-dimensional phase-space integrals \cite{Vogelsang:1990ug}.

As an important check on the correctness of our new NLO expressions, we rederive also the
HQ coefficient functions relevant for the unpolarized DIS structure functions
$F_2$ and $F_L$ and compare them to the known results in the literature \cite{Laenen:1992zk}.
To this end, we introduce a notation throughout the paper that covers both polarized and unpolarized
coefficient functions in a compact way.
In the polarized case, we shall compare the results of our full NLO calculation to the
above mentioned partial, analytical results, namely the partonic cross section
for $\Pggx$-light quark Compton scattering \cite{Buza:1996xr,Blumlein:2016xcy}
and the expressions in the limit $Q^2\gg m^2$ \cite{Buza:1996xr}.
We also take the limit $Q^2\to 0$ whenever possible to check our expressions
against the known results in the limit of photoproduction given in Ref.~\cite{Bojak:1998bd}.

At the partonic level, we will present numerical results for all virtual photon-parton cross sections 
at ${\cal{O}}(\alpha_s^2)$, i.e., the HQ scaling functions at NLO accuracy 
in the strong coupling $\alpha_s$, as a function of the available 
partonic center-of-mass system (c.m.s.) energy $s$ for various values of the ratio $\xi\equiv Q^2/m^2$.
Compact analytical expressions are given whenever possible, otherwise results are available upon request.
In addition, we compare numerically to the known results for both
polarized photoproduction \cite{Bojak:1998bd} and the unpolarized
HQ scaling functions. The latter are relevant for the computation of the DIS structure functions $F_2$ and
$F_L$ \cite{Laenen:1992zk,Riemersma:1994hv,Harris:1995tu}.
The behavior of the HQ scaling functions close to threshold is analyzed in some detail,
and analytical expressions are provided up to the subleading level.

As we have already mentioned, our results will be of particular relevance 
for studies of longitudinally polarized DIS at a future EIC, in particular,
for global QCD analyses aiming at a much improved extraction of the elusive gluon helicity 
density in the small $x$ regime.
As a first phenomenological application, we provide some numerical estimates 
for the charm contribution to $g_1$ and the experimentally relevant double-spin asymmetry $A_1$
in the kinematic domain accessible to a future EIC
and shall comment on their sensitivity to $\Delta \Pg(x,Q^2)$.
To get an idea of the theoretical uncertainties inherent to the polarized electroproduction of
HQ at NLO accuracy, we investigate the residual dependence of our results
on the unphysical factorization and renormalization scales as well as on the 
choice of the HQ mass $m$.
The obtained NLO expressions for HQ production in spin-dependent DIS will 
help to revisit and supplement existing studies of the expected impact of EIC data on
furthering our knowledge of the spin structure of nucleons in terms of helicity PDFs
\cite{Aschenauer:2012ve}.
Finally, as an outlook, we mention further extensions of our NLO calculations to more exclusive
HQ distributions and correlations \cite{ref:prep}, again, performed along the lines
of already existing computations for the corresponding unpolarized expressions
\cite{Laenen:1992xs,Harris:1995tu}.

The remainder of the paper is organized as follows: in Sec.~\ref{sec:tech} we 
present the necessary technical framework and notation adopted in our calculations.
We discuss the various $\Pggx$-gluon and $\Pggx$-light quark induced contributions to the
polarized electroproduction of HQs comprising real emission and virtual one-loop corrections 
and how the renormalization and mass factorization are performed. 
In Sec.~\ref{sec:partonic} we show numerical results for the polarized HQ scaling functions and 
discuss comparisons to various existing results in the literature. We also elaborate a
bit on the threshold and high energy limits of the partonic coefficient functions.
Some first phenomenological studies relevant for an EIC are presented in
Sec.~\ref{sec:pheno} including discussions of theoretical uncertainties due to
variations of the renormalization and factorization scales and the actual value
of the HQ mass used in the calculations. We summarize the main results in Sec.~\ref{sec:summary}
and present an outlook to related work in progress.
Finally, the two Appendices collect some analytic expressions that are too lengthy for the main
body of the paper.

\section{Technical Framework \label{sec:tech}}
\subsection{General Remarks, Kinematics, and Notation \label{sec:notation}}
We will study heavy flavor production in longitudinally polarized deep-inelastic 
lepton-nucleon scattering, more specifically, the NLO QCD corrections 
to the relevant, underlying virtual photon-parton scattering cross sections 
and, in particular, the corresponding, HQ contributions to the inclusive,
spin-dependent DIS structure function $g_1(x,Q^2)$.
We limit ourselves to the neutral current process mediated by the exchange of a virtual photon $\Pggx(Q^2)$,
where $Q^2\equiv -q^2\ll M_{\PZ}^2$ with $M_{\PZ}$ the mass of the $Z$ boson.

Therefore, we need to compute the partonic processes
\begin{equation}
\Pggx(q) + i(k_1) \rightarrow \PQ(p_1)+\PaQ(p_2)+j(k_2)\;,
\label{eq:proc}
\end{equation}
where $q$, $k_1$, $k_2$, $p_1$, and $p_2$ label the four-momenta of the
virtual photon scattering off a parton $i$ and producing a heavy quark-antiquark pair, $\PQ$ and $\PaQ$, and,
at NLO accuracy, up to one additional parton $j$. 
In (\ref{eq:proc}), $i$ and $j$ can be both either a gluon $\Pg$ or a light
(anti)quark $\Pq$ ($\Paq$). 
As is customary, we adopt the on-shell scheme for the mass of the HQs, 
hence $m$ denotes their pole mass, i.e., $p_1^2=p_2^2=m^2$.
In this paper, as in Ref.~\cite{Laenen:1992zk}, we are interested in the single-inclusive DIS cross section for a
detected heavy \textit{anti}quark $\PaQ(p_2)$, hence, all other final-state particles in Eq.~(\ref{eq:proc})
are integrated out. The corresponding cross section for an observed heavy quark $\PQ$ can be derived from our results
by some appropriate substitutions as we shall discuss later.

The usual kinematical variables for HQ electroproduction (\ref{eq:proc}) are
\begin{eqnarray}
s &=& (q+k_1)^2\,\nonumber\\
t_1 &=& t-m^2=(k_1-p_2)^2-m^2\,,\nonumber\\
u_1 &=& u - m^2 = (q-p_2)^2 -m^2 \;,
\label{eq:lo-vars}
\end{eqnarray}
and, for convenience, we also define  $s' \equiv s-q^2$ and $u_1' \equiv u_1 - q^2$. 
The relevant virtual photon-parton scattering cross sections schematically read
\begin{equation}
\label{eq:proj}
\frac{d^2\sigma_{k}(s,t_1,u_1,q^2,m^2)}{dt_1du_1} = \tilde b_k(n)\; \hat {\mathcal P}^{\Pgg}_{k,\mu\mu'}W^{\mu\mu'}\;,
\end{equation}
where $\tilde b_k(n)$ is the required normalization factor to be specified below.
The $d^2\sigma_{k}$ can derived from the helicity-dependent partonic tensor $W^{\mu\mu'}$, see,
for instance, Ref.~\cite{Vogelsang:1990ug,Vogelsang:1990ka},
by applying appropriate projection operators \cite{Laenen:1992zk,Vogelsang:1990ug}
$\hat {\mathcal P}^{\Pgg}_{k,\mu\mu'}$, $k\in\{G,L,P\}$:
\begin{align}
\nonumber
\hat {\mathcal P}^{\Pgg}_{G,\mu\mu'} &= -g_{\mu\mu'}\;, \\ 
\nonumber
\hat {\mathcal P}^{\Pgg}_{L,\mu\mu'} &= -\frac{4q^2}{{s'}^2} k_{1,\mu}k_{1,\mu'} \;,\\
\hat {\mathcal P}^{\Pgg}_{P,\mu\mu'} &= i\varepsilon_{\mu\mu'\rho\rho'}\frac{q^{\rho}k_1^{\rho'}}{s'}\;.
\label{eq:proj-op}
\end{align}
The Lorentz indices $\mu$ and $\mu'$ in (\ref{eq:proj}) and (\ref{eq:proj-op})
refer to the virtual photon originating from the scattered lepton $\Pl'$ in inclusive 
DIS $\Pl(l) + \PN(P) \to \Pl'(l') + X$, which determines both the momentum $q=l-l'$ 
exchanged by the virtual photon as well as the Bjorken
scaling variable $x=-q^2/(2P\cdot q)$ where $P$ is the four-momentum of the nucleon $\PN$.

The operators in Eq.~(\ref{eq:proj-op}) project onto the different partonic cross sections 
or Lorentz structures that appear in the partonic tensor $W^{\mu\mu'}$ when the helicity of 
the incoming parton $i$ in (\ref{eq:proc}) is not averaged over. 
$k=G$ and $k=L$ refer to the usual unpolarized cross sections for HQ electroproduction
that have been calculated up to NLO accuracy in Ref.~\cite{Laenen:1992zk}.
The related transverse ($k=T$) partonic cross section is then obtained by 
\begin{align}
\label{eq:def-trans}
d^2\sigma_T &= d^2\sigma_G + \tilde b_G(n)\,d^2\sigma_L\;.
\end{align}
For $k=P$ one projects onto the helicity-dependent partonic cross section
$d^2\sigma_P$ that appears in the anti-symmetric part of $W^{\mu\mu'}$,
for which we will compute the full NLO corrections for the first time
in this paper. We note that $d^2\sigma_P$ is usually denoted as
$d^2\Delta \sigma$ in the literature as it actually refers to the 
measuring the difference of parallel and antiparallel alignments of the
lepton and nucleon spins in DIS. 

Upon further integration over $t_1$ and $u_1$ 
the different projections $k$ in (\ref{eq:proj}) lead to the HQ contributions
to the customary unpolarized and polarized hadronic DIS structure functions, 
$F_{1,2,L}(x,Q^2)$ and $g_1(x,Q^2)$, respectively, which can be
measured in experiment and dependent only on the DIS variables $x$ and $Q^2$. 
More specifically, $k=T$, $L$, and $P$ 
correspond to $2x F_1$, $F_L$, and $2x g_1$, respectively, 
and $F_2=2x F_1 + F_L$.

To compute the double-differential and total $\Pggx$-parton cross sections in 
(\ref{eq:proj}) we choose to work in $n=4+\epsilon$ dimensions to regulate
the soft/infrared (IR), collinear/mass, and ultraviolet (UV) divergencies
in intermediate steps of our calculation. 
Unfortunately, dimensional regularization is known to be nontrivial for spin-dependent processes, 
as they involve the Dirac-matrix $\gamma_5$ and the Levi-Civita tensor $\varepsilon_{\mu\nu\rho\sigma}$ 
to project onto fermion and bosons states of definite helicity, respectively, see, for instance, 
Ref.~\cite{Craigie:1984tk}, which both are genuinely four-dimensional objects. 
To this end, we adopt the commonly used HVBM prescription \cite{tHooft:1972tcz} 
to define them, which leads to the nuisance of $(n-4)$-dimensional ''hat-momenta'' in phase-space calculations,
apart from appearance of the ordinary $n$-dimensional scalar products of momenta \cite{Vogelsang:1990ug}.

In addition to the projection (\ref{eq:proj-op}) onto the various Lorentz structures $k$ in 
$W^{\mu\mu'}$, one has to control also the helicity of the incoming parton $i$, 
gluon or (anti)quark, when computing $d^2\sigma_k$. 
The final-state spins in (\ref{eq:proc}) are always summed over, and the initial-state
helicities of the parton $i$ are averaged over if $k\in\{G,L\}$ but for $k=P$ 
we need to project onto the helicity difference that is probed by $d^2\sigma_P$. 
For incoming gluons this is achieved by \cite{Craigie:1984tk}
\begin{align}
\hat {\mathcal P}^{\Pg}_{G,\nu\nu'} = \hat {\mathcal P}^{\Pg}_{L,\nu\nu'} 
= -g_{\nu\nu'}\;,\;\;\; 
\hat {\mathcal P}^{\Pg}_{P,\nu\nu'} = 2i\varepsilon_{\nu\nu'\rho\rho'}\frac{k_1^{\rho}q^{\rho'}}{s'}
\label{eq:proj-gluon}
\end{align}
where $\nu$ and $\nu'$ refer to the Lorentz indices of the gluon.
By choosing just $-g_{\nu\nu'}$ for $k\in\{G,L\}$ instead of the full physical
polarization sum, we decided to include also contributions from incoming external
ghosts in our calculations to cancel unphysical polarizations of the gluon;
see, for instance, Ref.~\cite{Bojak:1998bd} on how this was handled in
the corresponding calculation of (un)polarized HQ photoproduction.

As all initial-state (anti)quarks in (\ref{eq:proc}) are taken as
massless partons, the relevant projection operators onto definitive helicity 
states are given by \cite{Craigie:1984tk}
\begin{align}
\nonumber
\hat {\mathcal P}^{\Pq}_{G,aa'} &= \hat {\mathcal P}^{\Pq}_{L,aa'} = \left(\slashed k_1\right)_{aa'}\;,\;\;
\hat {\mathcal P}^{\Pq}_{P,aa'} = -\left(\gamma_5\slashed k_1\right)_{aa'}\;,\\
\hat {\mathcal P}^{\Paq}_{G,bb'} &= \hat {\mathcal P}^{\Paq}_{L,bb'} = \left(\slashed k_1\right)_{bb'}\;,\;\;
\hat {\mathcal P}^{\Paq}_{P,bb'} = \left(\gamma_5\slashed k_1\right)_{bb'}\;,
\end{align}
where $a$ and $a'$ ($b$ and $b'$)
refer to the Dirac-index of the initial (anti)quark spinor in the relevant matrix elements given below.  
 
For completeness, the normalization factors $\tilde b_k(n=4+\epsilon)=b_k(\epsilon)$ in Eq.~(\ref{eq:proj}) are given by
\begin{align}
b_G(\epsilon) = \frac{1}{2+\epsilon}\;,\;\;
b_L(\epsilon) = b_P(\epsilon) = 1\;.
\label{eq:b-norm}
\end{align}

In the computations of the Feynman diagrams at NLO accuracy, 
the derivations of the necessary phase space integrals,
and for finding compact analytical expressions we have extensively
made use of the computer algebra program \texttt{Mathematica}~\cite{ref:mathematica}
and the packages \texttt{TRACER}~\cite{ref:tracer} and 
\texttt{HEPMath}~\cite{ref:hepmath}.
\subsection{Born Cross Section in $n$ Dimensions \label{sec:born}}
\begin{figure}[ht!]
\begin{center}
\includegraphics[width=0.25\textwidth]{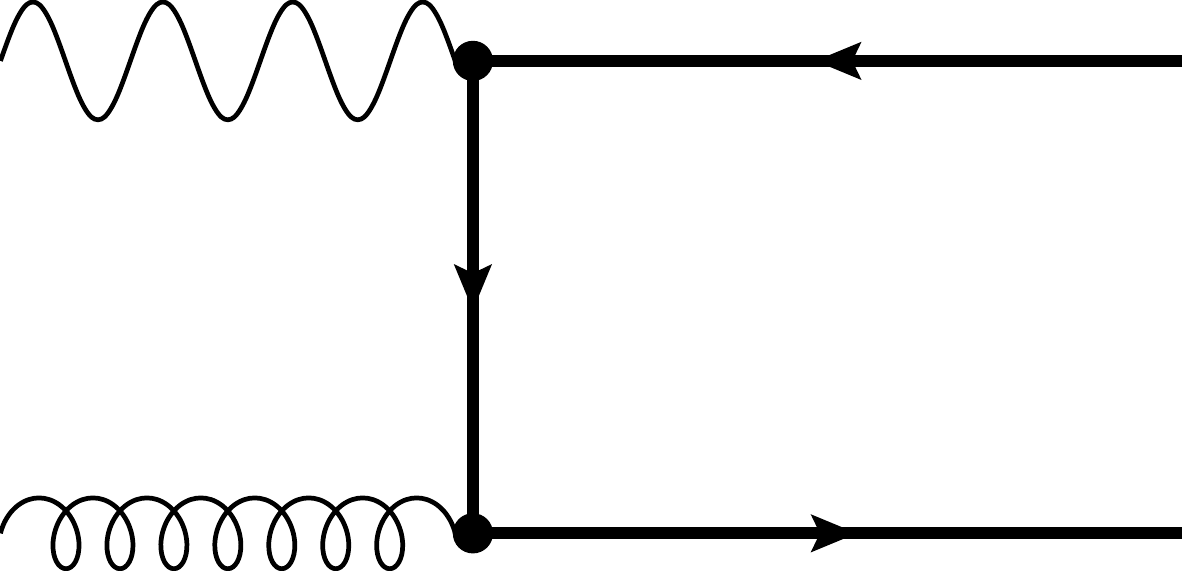}
\end{center}
\vspace*{-0.4cm}
\caption{PGF process $\Pggx\Pg\to \PQ\PaQ$ 
at LO accuracy. A second Feynman diagram (not shown) is obtained by
reversing the heavy quark lines.}\label{fig:pgf-born}
\end{figure}

For HQ electroproduction at LO accuracy in pQCD 
we only have to consider the photon-gluon-fusion (PGF) process, 
\begin{equation}
\Pggx(q) + \Pg(k_1) \rightarrow \PQ(p_1)+\PaQ(p_2)\;,
\label{eq:pgf-lo}
\end{equation}
depicted in Fig.~\ref{fig:pgf-born}, and where the four-momenta
are labeled by $q$, $k_1$, $p_1$, and $p_2$.

The relevant LO matrix elements for the PGF process
$\Md^{(0),j}_{\mu\nu}$, $j=1,2$, summed and squared, and 
properly projected onto the polarizations of both the photon and the gluon, see Sec.~\ref{sec:notation},
can be written as
\begin{align}
\nonumber
&\hat {\mathcal P}_{k}^{\Pgg,\mu\mu'}
\hat {\mathcal P}_{k}^{\Pg,\nu\nu'} \sum_{j,j'=1}^2\Md^{(0),j}_{\mu\nu}\left(\Md^{(0),j'}_{\mu'\nu'}\right)^* 
\nonumber \\
&= 8g^2\mu_D^{-\epsilon}e^2\,e_H^2\,N_C\,C_F\,B_{k,\tQED}\;.
\label{eq:me-born}
\end{align}
$g$ and $e$ denote the strong and electromagnetic coupling, respectively,
and $e_H$ is the charge of the heavy quark $\PQ$ in units of $e$. $N_C=3$ is the number
of colors and $C_F=(N_C^2-1)/(2N_C)$ the Casimir constant of the $SU(N_C)$ gauge group. We need to compute (\ref{eq:me-born}) in
$n=4+\epsilon$ dimensions as the full Born cross section will be needed, e.g., 
for mass factorization at NLO accuracy.
Hence, an arbitrary mass scale $\mu_D$ is introduced in Eq.~(\ref{eq:me-born}) to keep the
strong coupling dimensionless in $n$ dimensions.
The quantities $B_{k,\tQED}$ in Eq.~(\ref{eq:me-born}) represent the QED analogues
of the LO PGF process (\ref{eq:pgf-lo}) for each projection $k$ and, up to $\mathcal{O}(\epsilon^2)$, 
are given by
\begin{align}
B_{G,\tQED} &= \frac{t_1}{u_1} + \frac{u_1}{t_1} + \frac{4m^2s'}{t_1u_1}\left(1-\frac{m^2s'}{t_1u_1}\right)
+\frac{2s'q^2}{t_1u_1} \nonumber \\
&+\frac{2q^4}{t_1u_1} + \frac{2m^2q^2}{t_1u_1}\left(2-\frac{{s'}^2}{t_1u_1}\right) \nonumber \\
&+\epsilon\left\{ -1 + \frac{{s'}^2}{t_1u_1} + \frac{s'q^2}{t_1u_1} -
\frac{q^4}{t_1u_1} - \frac{m^2q^2{s'}^2}{t_1^2u_1^2} \right\} \nonumber \\
&+ \epsilon^2\frac{{s'}^2}{4t_1u_1}\;,\label{eq:BQEDG}  \\
B_{L,\tQED} &= -\frac{4q^2}{s'}\left(\frac s {s'} - \frac{m^2s'}{t_1u_1}\right)\;,  \label{eq:BQEDL}\\
B_{P,\tQED} &= \frac 1 2\left(\frac{t_1}{u_1}+\frac{u_1}{t_1}\right)
\left(\frac{2m^2 s'}{t_1u_1}-1 - \frac{2q^2}{s'}\right)\;.
\label{eq:BQEDP}
\end{align}

The required $n$-dimensional phase space $d\text{PS}_2$ is straightforwardly obtained in
the center-of-mass system (c.m.s) of the produced HQ pair, i.e.,
$p_1+p_2=q+k_1=(\sqrt s,\vec 0)$, utilizing the mass-shell conditions $p_1^2 = p_2^2=m^2$,
and reads \cite{Laenen:1992zk,Bojak:1998bd}
\begin{align}
d\text{PS}_2 &= \frac {2\pi S_\epsilon}{s'\Gamma[(n-2)/2]}\, \delta(s'+t_1+u_1)
\nonumber \\
&\times \left(\frac{(t_1u_1'-s'm^2)s' - q^2t_1^2}{s'^2}\right)^{(n-4)/2}\,dt_1du_1\nonumber \\
\label{eq:dps2}
&\equiv h_2(n)\,\delta(s'+t_1+u_1)\, dt_1 du_1\;,
\end{align}
where the Gamma function is represented by $\Gamma$ and $S_\epsilon = (4\pi)^{(-n/2)}$. 
Combining $d\text{PS}_2$ with the matrix element squared in Eq.~(\ref{eq:me-born}) and the necessary
prefactors, i.e. spin average and flux factor, the double-differential partonic cross section for the PGF process at LO in $n=4+\epsilon$
dimensions can be written as
\begin{align}
\label{eq:born-ndim}
&{s'}^2\frac{d^2\sigma_{k,\Pg}^{(0)}}{dt_1du_1} = \alpha\,\alpha_s\, K_{\Pg\Pgg}
b_k(\epsilon) \frac{2^6\pi^3S_\epsilon}{\Gamma(1+\epsilon/2)} 
\nonumber \\
&\times E_k(\epsilon) 
\left(\frac{\mu_D^2}{m^2}\right)^{-\epsilon/2}
\left(\frac{(t_1u_1'-s'm^2)s' - q^2t_1^2}{m^2{s'}^2}\right)^{\epsilon/2}
\nonumber \\
&\times e_H^2\, N_C\,C_F\, B_{k,\tQED}\,\delta(s'+t_1+u_1)\;.
\end{align}
Note that here and in what follows, we suppress the arguments of $d^2\sigma_{k,\Pg}^{(0)}$
on the left-hand-side (l.h.s.) of Eq.~(\ref{eq:born-ndim}) unless indicated otherwise.
The color average for the incoming gluon is given by $K_{\Pg\Pgg} = 1/(N_C^2-1)$,
\begin{equation}
E_G(\epsilon) = E_L(\epsilon) = \frac 1 {1+\epsilon/2}\;,\;\;\; E_P(\epsilon) = 1\;,
\end{equation}
properly accounts for additional degrees of freedom in $n$ dimensions for initial-state bosons,
and $\alpha\equiv e^2/(4\pi)$ and $\alpha_s \equiv g^2/(4\pi)$.
From Eq.~(\ref{eq:born-ndim}) one can easily obtain analytical expressions for
the total partonic cross section, i.e., integrated over $dt_1 du_1$,
and its threshold limit $s\to 4m^2$, for each projection $k$ at LO accuracy.
We shall get back to this point in Sec.~{\ref{sec:partonic} when we have derived also
the complete NLO corrections to HQ electroproduction. 

\subsection{One-Loop Virtual Corrections \label{sec:virtual}}
\begin{figure*}[th!]
\begin{center}
\includegraphics[width=0.80\textwidth]{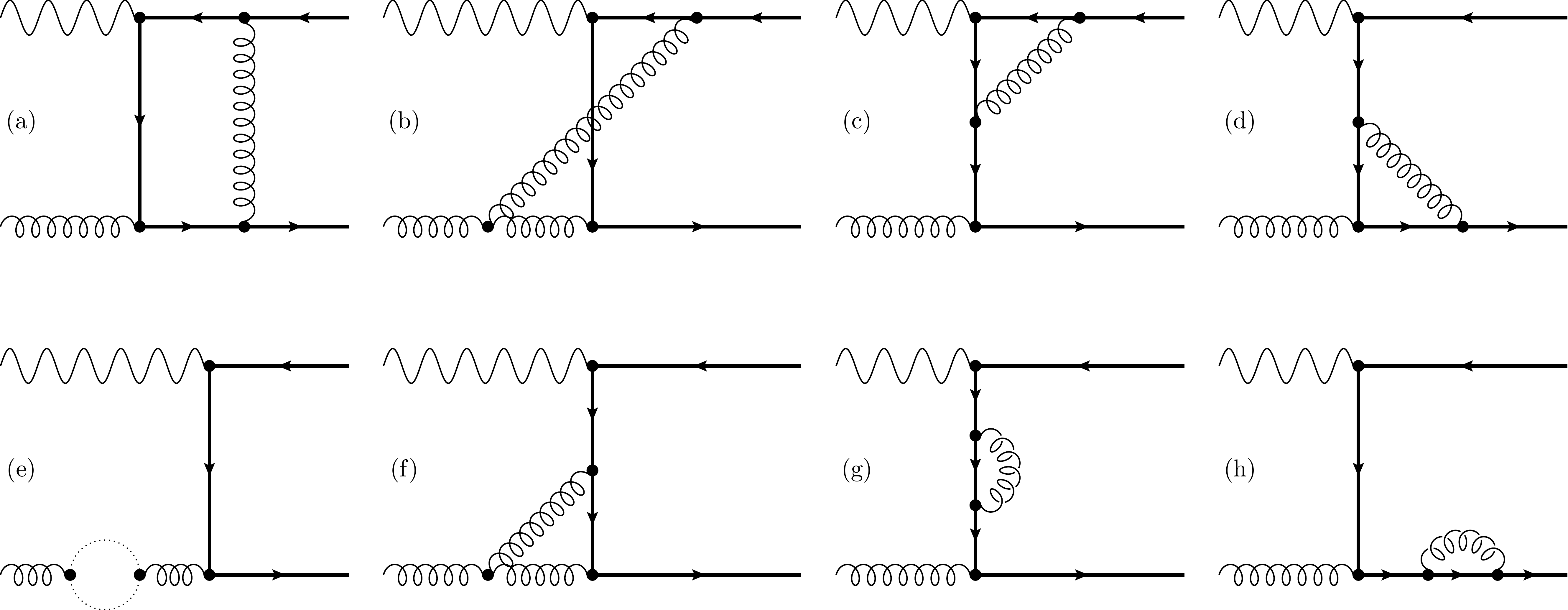}
\end{center}
\vspace*{-0.4cm}
\caption{Examples for Feynman diagrams contributing to the one-loop virtual corrections
to the PGF process. The dotted line in the gluon self-energy correction (e) 
can represent either a quark, gluon, or ghost loop. Remaining diagrams are obtained by appropriate crossing.}
\label{fig:feyn-virt}
\end{figure*}
The one-loop virtual corrections to the PGF process are displayed in
Fig.~\ref{fig:feyn-virt}. At NLO accuracy only all possible interferences of
the two Born diagrams $\Md^{(0)}_{j',\mu'\nu'}$
with the one-loop virtual amplitudes $\Md^{(1),V}_{j,\mu\nu}$ contribute.
It is customary in HQ photo- and electroproduction to organize the results into
the Abelian QED and non-Abelian OK parts \cite{Smith:1991pw,Bojak:1998bd,Laenen:1992zk},
i.e.\ their color structure. 
The contribution at NLO, summed over all amplitudes and
properly projected onto the polarizations of both the photon and the gluon, 
can be written as
\begin{align}
\nonumber
&\!\!\hat {\mathcal P}_{k}^{\Pgg,\mu\mu'}
\hat {\mathcal P}_{k}^{\Pg,\nu\nu'}  
\sum_{j,j'} 2\,\text{Re} \left[\Md^{(1),V}_{j,\mu\nu}\left(\Md^{(0)}_{j',\mu'\nu'}\right)^*\right]
\nonumber\\
&\!\!\!\!=8g^4\mu_D^{-\epsilon}  e^2\,e_H^2\, N_C\, C_F\,C_\epsilon
\left( C_A V_{k,\tOK} + 2C_F V_{k,\tQED}\right)
\label{eq:me-virt}
\end{align}
where
\begin{equation}
C_\epsilon = \frac{1}{16\,\pi^2}\exp\left(\frac{\epsilon}{2}\left[\gamma_E-\ln(4\pi)\right]\right)
\label{eq:ceps}
\end{equation}
with $\gamma_E$ the Euler-Mascheroni constant.

As in \cite{Smith:1991pw,Bojak:1998bd,Laenen:1992zk} the computation of the one-loop virtual 
amplitudes in (\ref{eq:me-virt}) proceeds as follows: all divergencies are
regulated in $n=4+\epsilon$ dimensions, internal gluon propagators are calculated in
Feynman gauge, light quark masses are all put to zero,
and the tensorial loop integrals are reduced to scalar ones using an adapted
Passarino-Veltman decomposition method \cite{Passarino:1978jh}, which is described in detail in 
Ref.~\cite{Bojak:2000eu,Bojak:1998bd}.
The required one-loop scalar one-, two-, three-, four-point functions can be all found in the literature.
For instance, App.~A of Ref.~\cite{Beenakker:1988bq} collects all well-tested scalar integrals without dependence on the virtual 
photon momentum $q$, and those that carry a $q$-dependence are listed in App.~A of Ref.~\cite{Laenen:1992zk}.
We have performed extensive checks of the latter set of integrals both analytically and with the help of
\texttt{LoopTools} \cite{Hahn:1998yk} and fully agree with the results given in \cite{Laenen:1992zk} except for
the four-point function $D_0$ with three massive propagators. Here we find, using the notation of \texttt{LoopTools}
for the arguments of $D_0$ 
\begin{align}
\nonumber
&D_0(m^2,0,q^2,m^2,t,s,0,m^2,m^2,m^2) = \frac{i C_\epsilon}{\beta s t_1}\\
\nonumber
&\times \Bigg[ -\frac{2}{\epsilon} \ln(\chi) -2\ln(\chi)\ln\left(\frac{-t_1}{m^2}\right)
+\DiLog(1-\chi^2)-4\zeta(2)\\
\nonumber
&+\ln^2(\chi_q) + 2\DiLog(-\chi\chi_q)+2\DiLog\left(\frac{-\chi}{\chi_q}\right)+2\ln(\chi\chi_q)\\
&\times \ln(1+\chi\chi_q)+2\ln\left(\frac{\chi}{\chi_q}\right)\ln\left(1+\frac{\chi}{\chi_q}\right) \Bigg]\;,
\label{eq:d0}
\end{align}
which also agrees with Box~16 in Ref.~\cite{Ellis:2007qk}.
Here, we have adopted a set of additional partonic variables 
\begin{align}\label{eq:partonic-var}
\nonumber
0\leq \rho = \frac {4m^2} s\leq 1\;&,\;\;     \rho_q = \frac {4m^2} {q^2}\leq 0\;,\\ 
\nonumber
0\leq \beta = \sqrt{1-\rho}\leq 1\;&,\;\;     1\leq \beta_q = \sqrt{1-\rho_q}\;,\\
0\leq \chi = \frac{1-\beta}{1+\beta}\leq 1 \;&,\;\; 0\leq \chi_q = \frac{\beta_q-1}{\beta_q+1}\leq 1
\end{align}
that will be extensively used in the following in order to optimize the analytical expressions.
In order to have $0\leq \chi_q \leq 1$, we had to introduce an additional minus sign 
into the definition of $\chi_q$ as compared to that for $\chi$.

The complicated structure of (\ref{eq:d0}) already suggests that it is an impossible task 
to give compact analytical expressions for $V_{k,\tOK}$ and $V_{k,\tQED}$ in (\ref{eq:me-virt}).
Here, we quote only their singular parts that, as they should, are proportional to the 
Born result for each projection $k\in\{G,L,P\}$:
\begin{align}
\label{eq:vok}
V_{k,\tOK} &= -2B_{k,\tQED} \Bigg\{ \frac 4 {\epsilon^2} + \frac{2}{\epsilon} 
\Bigg[ \ln\left(\frac{-t_1}{m^2}\right) 
+ \ln\left(\frac{-u_1}{m^2}\right)\nonumber\\
&+\frac{s-2m^2}{s\beta}\ln(\chi)\Bigg] \Bigg\} + \mathcal{O}(\epsilon^0)\;,\\
V_{k,\tQED} &= -2B_{k,\tQED}\left[ 
1-\frac{s-2m^2}{s\beta}\ln(\chi)\right] \frac 2 \epsilon + \mathcal{O}(\epsilon^0)\;.
\label{eq:vqed}
\end{align}
The double poles in $V_{k,\tOK}$ originate from diagrams where
soft and collinear singularities can coincide.

Collecting all prefactors, the bare 
double differential partonic one-loop virtual PGF cross section can be written as
\begin{align}
\nonumber
& {s'}^2\frac{d^2\sigma_{k,\Pg}^{(1),V}}{dt_1du_1}\Bigg|_{\text{bare}} = 
\alpha\alpha_s^2  K_{\Pg\Pgg} b_k(\epsilon) \frac{2^8\pi^4 S_\epsilon}{\Gamma(1+\epsilon/2)} 
\\
\nonumber
& \delta(s'+t_1+u_1) C_\epsilon\left(\frac{\mu_D^2}{m^2}\right)^{-\epsilon/2} 
\left(\frac{(t_1u_1'-s'm^2)s' - q^2t_1^2}{m^2{s'}^2}\right)^{\epsilon/2}\nonumber\\
&E_k(\epsilon) N_C\,C_F\, e_H^2
\left( C_A V_{k,\tOK} + 2C_F V_{k,\tQED}\right)\;.
\label{eq:virt-bare}
\end{align}
All UV divergencies in the one-loop amplitude (\ref{eq:virt-bare}) are removed by mass and coupling 
constant renormalization for which we choose the same, modified $\overline{\text{MS}}$ prescription 
as in Refs.~\cite{Smith:1991pw,Bojak:1998bd,Laenen:1992zk}.
Specifically, the heavy (anti)quark is renormalized on-shell, and the HQ masses $m$ are defined
as pole masses. Note that the self-energies on external legs are not included yet in the results
given in Eqs.~(\ref{eq:vok}) and (\ref{eq:vqed}). 
The strong coupling is renormalized in such a way that the HQ loop to the gluon self-energy,
shown in Fig.~\ref{fig:feyn-virt}~(e), is removed. This leads to a fixed-flavor number scheme with
$n_{lf}=n_f-1$ light quark flavors active in the running of $\alpha_s$ and the evolution of the PDFs. 
Hence, the UV-renormalization of the bare, one-loop partonic cross section 
at a renormalization scale $\mu_R$ is achieved by,
see Refs.~\cite{Smith:1991pw,Bojak:1998bd,Laenen:1992zk,Nason:1989zy} for further details,
\begin{align}
\nonumber
&\frac{d^2\sigma_{k,\Pg}^{(1),V}}{dt_1du_1}
=\frac{d^2\sigma_{k,\Pg}^{(1),V}}{dt_1du_1}\Bigg|_{\text{bare}} 
+ 4\pi\alpha_s(\mu_R^2)\,C_\epsilon\,\left(\frac{\mu_D^2}{m^2}\right)^{-\epsilon/2}\\
&\times \left\{\left[\frac 2 \epsilon +\ln\left(\frac{\mu_R^2}{m^2}\right)\right]\beta_0^f 
+\frac 2 3 \ln\left(\frac{\mu_R^2}{m^2}\right)\right\}\frac{d^2\sigma_{k,\Pg}^{(0)}}{dt_1du_1}\;.
\end{align}
Here, the first term in the square bracket corresponds to the usual $\overline{\text{MS}}$ scheme
and the second term removes the HQ loop from the gluon self-energy. 
$\beta_0^f = (11C_A- 2n_{f})/3$ is the first order coefficient of the QCD beta function.
In what follows, we will often drop the scale in the strong coupling, i.e., $\alpha_s$ 
has to be understood as $\alpha_s(\mu_R^2)$.

\subsection{Single Gluon Radiation Corrections \label{sec:realcorr}}
Apart from the virtual corrections to the Born PGF process considered
in the previous subsection, we also
need to compute at NLO accuracy the $\mathcal{O}(\alpha_s^2)$ corrections from
real gluon emission, i.e., the $2\to 3$ process
\begin{equation}
\Pggx(q) + \Pg(k_1) \rightarrow \PQ(p_1)+\PaQ(p_2) + \Pg(k_2)\;.
\label{eq:gluon23proc}
\end{equation}
A selection of contributing Feynman diagrams is depicted in Fig.~\ref{fig:feyn-gluon}.
\begin{figure}[ht!]
\begin{center}
\includegraphics[width=0.42\textwidth]{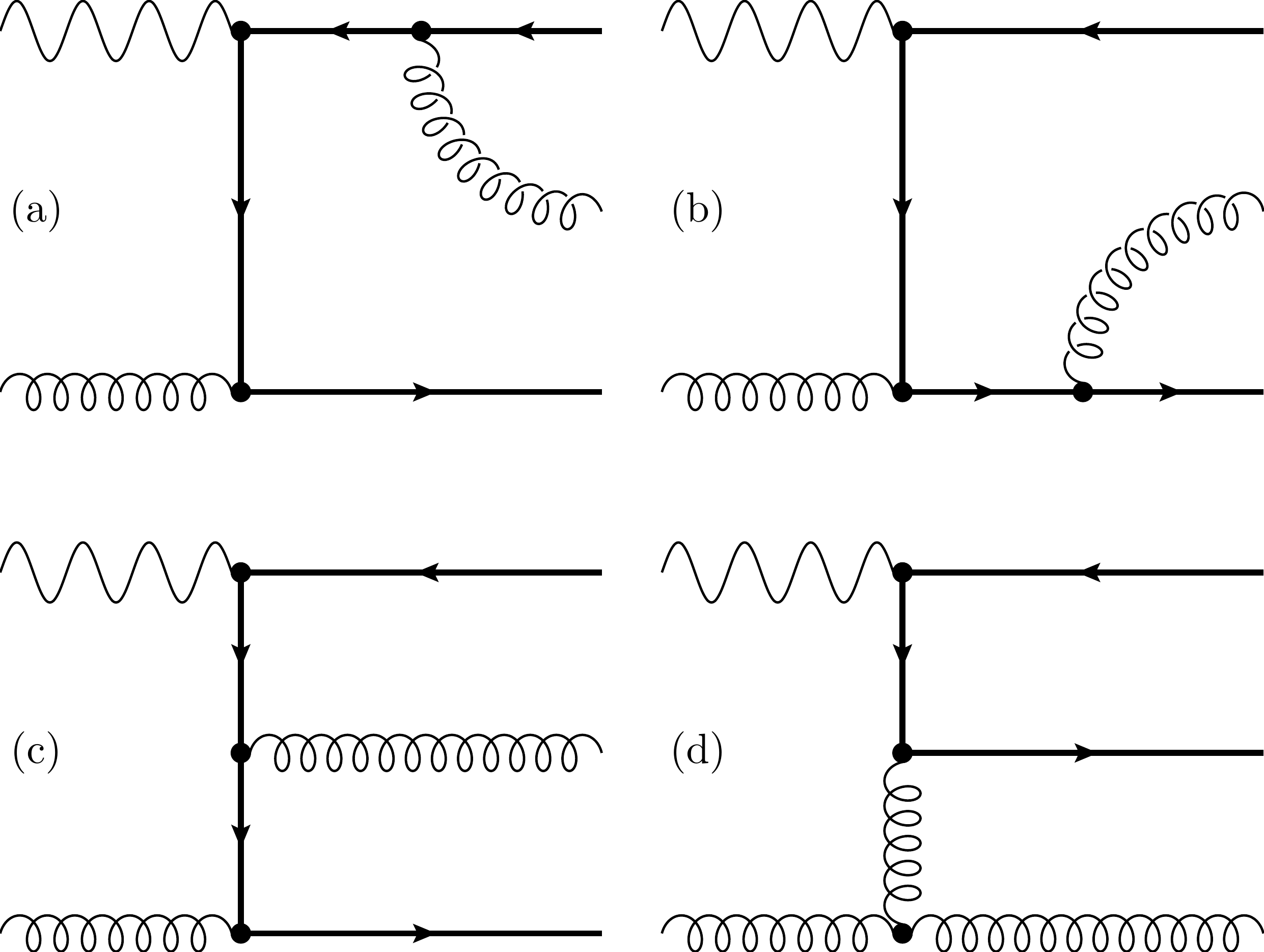}
\end{center}
\vspace*{-0.1cm}
\caption{Selected Feynman diagrams contributing to the real gluon emission
process in (\ref{eq:gluon23proc}). Remaining diagrams are obtained by appropriate crossing.}\label{fig:feyn-gluon}
\end{figure}

As before, we split the results according to their color structure into
QED and OK parts \cite{Smith:1991pw,Bojak:1998bd,Laenen:1992zk}. 
Only the latter contribution will develop singularities from
collinear gluon emissions originating from diagram (d) in Fig.~\ref{fig:feyn-gluon}
as we shall see below.
The relevant NLO matrix element squared for (\ref{eq:gluon23proc}), 
properly projected onto the polarizations of both the photon and the gluon, 
can be written as
\begin{eqnarray}
\nonumber
\hat {\mathcal P}_{k}^{\Pgg,\mu\mu'}
\hat {\mathcal P}_{k}^{\Pg,\nu\nu'}&& \!\!\!\!\!
\sum_{j,j'}\Md^{(1),\Pg}_{j,\mu\nu}\left(\Md^{(1),\Pg}_{j',\mu'\nu'}\right)^* = 8g^4\mu_D^{-2\epsilon}\,e^2\,e_H^2 \\
&\times& N_C\,C_F \left[ C_A\, R_{k,\tOK} + 2\,C_F\, R_{k,\tQED}\right].
\label{eq:pgf-me}
\end{eqnarray}
Apart from the Mandelstam variables $s,t_1$, and $u_1$ already used for the Born and virtual contributions,
it is convenient to introduce in addition
\begin{align}
\nonumber
s_3 &= (k_2+p_2)^2-m^2\;, & s_4 &=(k_2+p_1)^2-m^2\;,\\
\nonumber
s_5 &= (p_1+p_2)^2=-u_5\;, & t' &= (k_1-k_2)^2\;,\\
\nonumber
u' &= (q-k_2)^2\;, &         u_6 &=(k_1-p_1)^2-m^2\;, \\
u_7 &=(q-p_1)^2-m^2\;,
\end{align}
out of which only five are independent due to momentum conservation $k_1+q=p_1+p_2+k_2$.

The derivation of the $2\rightarrow 3$ phase space $d\text{PS}_3$ in
$n=4+\epsilon$ dimensions with two equal masses $m$ is standard \cite{Beenakker:1988bq,Bojak:1998bd} 
but some extra care is needed in our case for $k=P$ since the matrix element
squared in (\ref{eq:pgf-me}) will depend on $(n-4)$--dimensional scalar products
of momenta (usually labeled as ''hat momenta''), see, e.g., Ref.~\cite{Bojak:1998bd}. 
The momenta in (\ref{eq:gluon23proc}) are most conveniently parametrized
in the c.m.s.\ frame of the two outgoing, unobserved partons \cite{Gottfried:1964zz}, 
i.e., in our case, where $\PaQ$ will be observed, $p_1+k_2 = (\sqrt{s_4+m^2},\vec 0)$.
Then, only one $(n-4)$--dimensional scalar product, say $\hat{p}_1^2$, remains in the matrix elements squared, and 
the phase space calculations can be organized in such a way, that
the additional integration over the hat momenta space yields unity whenever hat
momenta are not present in (\ref{eq:pgf-me}), i.e., for $k=\{G,L\}$ and most of the
terms for $k=P$. We use
\begin{align}
\nonumber
d\text{PS}_3 &= \frac 1 {(4\pi)^n\Gamma(n-3)s'} \frac{s_4^{n-3}}{(s_4+m^2)^{n/2-1}} \\
\nonumber
&\left(\frac{(t_1u_1'-s'm^2)s' - q^2t_1^2}{s'^2}\right)^{(n-4)/2}\! dt_1 \,du_1 \, d\Omega_n \,d\hat{\mathcal I}\\
&\equiv h_3(n)\,dt_1\, du_1\, d\Omega_n\, d\hat{\mathcal I}\;,
\label{eq:dps3}
\end{align}
where $d\Omega_n = \sin^{n-3}(\theta_1)\,d\theta_1\sin^{n-4}(\theta_2)\,d\theta_2$ and
\begin{align}
d\hat{\mathcal I} &= \frac 1 {B[1/2,(n-4)/2]}\frac{x^{(n-6)/2}}{\sqrt{1-x}}dx 
\end{align}
with $x \equiv \hat p_1^2/\hat p_{1,\max}^2$, the Euler Beta function $B[a,b]$, and 
\begin{align}
\hat p_{1,\max}^2 &= \frac{s_4^2}{4(s_4+m^2)}\sin^2(\theta_1)\sin^2(\theta_2)\;.
\end{align}
The integration over the hat momenta space $d\hat{\mathcal I}$ yields only two possible results,
depending on whether or not the term to be integrated in the matrix element squared is proportional
to $\hat{p}_1^2$, i.e.,
\begin{equation}
\int\!d\hat{\mathcal I}\,\hat p_1^2 = \epsilon\, \hat p_{1,\max}^2 + {\mathcal O}(\epsilon^2)\;\;\text{or}\;\;
\int\!d\hat{\mathcal I}\, 1 = 1 \;.
\end{equation}
As it should, the contribution from the $\hat{p}_1^2$ integration is of ${\mathcal O}(\epsilon)$.
Nevertheless, when accompanied by collinear divergent 
angular integrals $\propto 1/{{t'}}$ in Eq.~(\ref{eq:pgf-me}) they will contribute to the final result for $k=P$. 
For the discussions below, it is instructive to compare the $n$-dimensional $2\to 2$ and $2\to 3$ phase space
factors $h_2(n)$ and $h_3(n)$ defined in Eq.~(\ref{eq:dps2}) and (\ref{eq:dps3}), respectively. One finds
\begin{align}
\nonumber
\frac{h_3(4+\epsilon)}{h_2(4+\epsilon)} &= \frac{S_\epsilon}{2\pi} 
\frac{\Gamma(1+\epsilon/2)}{\Gamma(1+\epsilon)} \frac{s_4^{1+\epsilon}}{(s_4+m^2)^{1+\epsilon/2}}\\
&= \frac{C_\epsilon}{2\pi}\left[1-\frac 3 8 \zeta(2)\epsilon^2\right]\frac{s_4^{1+\epsilon}}{(s_4+m^2)^{1+\epsilon/2}} 
+ \mathcal{O}(\epsilon^3),
\label{eq:ps23-diff}
\end{align}
where $\zeta$ denotes the Riemann Zeta function.

Putting everything together, the double differential partonic PGF cross section with one additional
real (R) gluon emission reads
\begin{align}
\nonumber
&{s'}^2\frac{d^2\sigma_{k,\Pg}^{(1),R}}{dt_1du_1} 
= \alpha\alpha_s^2  K_{\Pg\Pgg}  b_k(\epsilon)
\frac{2^7\pi^3 S_\epsilon^2}{\Gamma(1+\epsilon)}
\left(\frac{\mu_D^2}{m^2}\right)^{-\epsilon}  \frac{s_4}{s_4+m^2}   \\
\nonumber
&   
\left(\frac{(t_1u_1'-s'm^2)s' - q^2t_1^2}{m^2{s'}^2}\right)^{\epsilon/2} 
 \left(\frac{s_4^2}{m^2(s_4+m^2)}\right)^{\epsilon/2} 
\nonumber\\
& E_k(\epsilon)\,N_C\,C_F\, e_H^2\int\!d\Omega_n d\hat{\mathcal I}\,\left(C_A R_{k,\tOK} + 2C_F R_{k,\tQED}\right)\;.
\label{eq:xsec-gluonreal}
\end{align}
Analytical results for the phase space integrals in $\theta_1$ and $\theta_2$ appearing in Eq.~(\ref{eq:xsec-gluonreal})
are conveniently tabulated in Refs.~\cite{Beenakker:1988bq} and \cite{Bojak:1998bd}. The few additional
integrals originating from the extra powers $\sin^2(\theta_1)\sin^2(\theta_2)$ in  
$\int\!d\hat{\mathcal I}\,\hat p_1^2$ are straightforward to evaluate.
Following most previous calculations of HQ production \cite{Beenakker:1988bq,Smith:1991pw,Laenen:1992zk,Bojak:1998bd}, 
we proceed by splitting the real emission cross section $d^2\sigma_{k,\Pg}^{(1),R}$ 
into a hard (H) and soft (S) gluon radiation part based on $s_4>\Delta$ and $s_4\le \Delta$, respectively.
The auxiliary parameter $\Delta$ is chosen small enough to be negligible in comparison with all
kinematic quantities $s'$, $t_1$, $u_1$, and $m^2$; a typical choice being $\Delta/m^2\simeq 10^{-6}$.
In the hard regime, $\Delta$ effectively cuts off all IR singularities, and only collinear singularities remain. The
soft part will be combined with the virtual corrections to the PGF process computed in the previous subsection.

In general, the analytical results after phase space integration are way too lengthy to report here, except for the
collinear and soft limits of Eq.~(\ref{eq:xsec-gluonreal}).
Only the non-Abelian OK part contains mass singularities originating from collinear gluon splittings in
diagram (d) in Fig.~\ref{fig:feyn-gluon}, i.e., from terms proportional to
$1/{t'}$ in $R_{k,\tOK}$. They yield
\begin{eqnarray}
\nonumber
&&\lefteqn{\frac{s_4}{4\pi(s_4+m^2)}\int\!d\Omega_n d\hat{\mathcal I}\,C_A R_{k,\tOK} =} \\
&&-\frac 1 {u_1}B_{k,\tQED}\left(\!\begin{array}{l}s'\rightarrow x_1s'\\t_1\rightarrow x_1 t_1\end{array}\!\!\right) 
P^{k,(0)}_{\Pg\Pg,H}(x_1)\, \frac 2 \epsilon + {\mathcal O}(\epsilon^0) 
\label{eq:gluon-coll}
\end{eqnarray}
with $x_1 = -u_1/(s'+t_1)$ and, depending on the projection $k$, the appropriate hard part of the LO 
gluon-gluon splitting function \cite{Altarelli:1977zs}
\begin{align}
P^{G,(0)}_{\Pg\Pg,H}(x) &= C_A\left[\frac 2 {1-x} + \frac 2 x - 4 + 2x - 2x^2\right]\;,\\
P^{L,(0)}_{\Pg\Pg,H}(x) &= P^{G,(0)}_{\Pg\Pg,H}(x)\;,\\
P^{P,(0)}_{\Pg\Pg,H}(x) &= C_A\left[\frac 2 {1-x} - 4x + 2\right]\;.
\label{eq:pgg-hard}
\end{align}
The hard Abelian QED part is finite.

The soft gluon limit $k_2\to 0$, i.e., $s_4, s_3,$ and $t'\to 0$, of the matrix element squared (\ref{eq:pgf-me}) 
is readily derived and reads
\begin{eqnarray}
\nonumber
&&\lefteqn{\lim_{k_2\to 0}  \left(C_A\, R_{k,\tOK} + 2C_F\, R_{k,\tQED}\right) =} \\
&&\left(C_A S_{k,OK} + 2C_F S_{k,QED}\right) + \mathcal{O}(1/s_4,1/s_3,1/t')\;,\;
\end{eqnarray}
where
\begin{align}
\nonumber
S_{k,\tOK}  &= 2\left(\frac{t_1}{t's_3} + \frac{u_1}{t's_4}-\frac{s-2m^2}{s_3s_4}\right)B_{k,\tQED}\;,\\
S_{k,\tQED} &= 2\left(\frac{s-2m^2}{s_3s_4} - \frac{m^2}{s_3^2} - \frac{m^2}{s_4^2}\right)B_{k,\tQED}\;.
\label{eq:soft-me}
\end{align}
Note that the eikonal factors multiplying the Born PGF cross section $B_{k,\tQED}$ in (\ref{eq:soft-me})
neither depend on the photon's virtuality $q^2$ nor on the spin projection $k$. 

The phase space slicing introduced above allows one to perform not only the angular but also the
$s_4$ integrations in the soft limit of Eq.~(\ref{eq:pgf-me}) analytically. Since $d\text{PS}_3$ behaves as
$s_4^{1+\epsilon}$ in the limit $s_4\to0$, let us consider a generic function $\mathcal{H}(s_4)$ with
a soft pole $s_4^{-1+\epsilon}\mathcal S(s_4)$ and a finite part $\mathcal F(s_4)$,
to illustrate this point a bit further \cite{Bojak:1998bd}. With the help of the identity 
\begin{equation}
s_4^{-1+\epsilon} = \frac{\Delta^{\epsilon}}{\epsilon} \delta(s_4) + [s_4^{-1+\epsilon}]_{\Delta}\;,
\end{equation}
which is completely analogous to the corresponding one for the well-known ``$+$--distribution'', one easily
derives for small enough $\Delta$ that
\begin{align}
\nonumber
\int\limits_0^{s_{4,\max}} \!\!\mathcal{H} (s_4) 
&= \int\limits_0^{s_{4,\max}} \!\!\left[s_4^{-1+\epsilon}\mathcal S(s_4) + \mathcal F(s_4)\right]\\
&\simeq \frac{\Delta^\epsilon}{\epsilon}\mathcal S(0) + \int\limits_\Delta^{s_{4,\max}}\!\!\mathcal{H}(s_4)
\label{eq:soft-identity}
\end{align}
where $s_{4,\max}$ denotes the upper kinematic limit of the $s_4$ integration; see also Sec.~\ref{sec:partonic} below.

According to Eq.~(\ref{eq:soft-identity}), we can thus perform all angular integrations in the soft limit (\ref{eq:soft-me})
analytically and obtain for the QED and OK parts
\begin{widetext}
\begin{align}
\nonumber
\lim_{s_4\to 0}\,&\frac{s_4}{2\pi(s_4+m^2)} \left[1-\frac 3 8 \zeta(2)\epsilon^2\right] \frac{s_4}{\epsilon}
\int\!d\Omega_n \,d\hat{\mathcal I}\,S_{k,\tOK} \\
\nonumber
&=2B_{k,\tQED}\left\{ \frac 4 {\epsilon^2} + \frac 2 \epsilon \left[\ln\left(\frac{t_1}{u_1}\right)
+\frac{s-2m^2}{s\beta} \ln(\chi)\right]
-\ln^2(\chi) - \frac 3 2 \zeta(2)+\frac 1 2\ln^2\left(\frac{u_1\chi}{t_1}\right)\right. \\
&\left. \hspace{50pt} +\DiLog\left(1- \frac{t_1}{u_1\chi}\right) 
- \DiLog\left(1-\frac{u_1}{t_1\chi}\right) + \frac{s-2m^2}{s\beta}\left[
\DiLog(1-\chi^2)+\ln^2(\chi)
\right] \right\} + \mathcal O(\epsilon) \;,
\label{eq:soft-ok}
\end{align}
and
\begin{align}
\lim_{s_4\to 0}\,&\frac{s_4}{2\pi(s_4+m^2)} \left[1-\frac 3 8 \zeta(2)\epsilon^2\right] \frac{s_4}{\epsilon}
\int\!d\Omega_n \,d\hat{\mathcal I}\,S_{k,\tQED}\nonumber\\
&= 2B_{k,\tQED}  \left\{-\frac 2 \epsilon  \left(1+\frac{s-2m^2}{s\beta}\ln(\chi)\right) +1 
- \frac{s-2m^2}{s\beta} \left[ \ln(\chi)\left[1+\ln(\chi)\right] + \DiLog(1-\chi^2) \right] \right\} + \mathcal O(\epsilon)
\label{eq:soft-qed}
\end{align}
\end{widetext}
respectively.
The additional factors on the l.h.s.~of Eqs.~(\ref{eq:soft-qed}) and (\ref{eq:soft-ok})
originate from the difference between $h_3(n)$ and $h_2(n)$ given in (\ref{eq:ps23-diff})
since the limit $s_4\to 0$ implies the use of $2\to2$ kinematics.
For $k=\{G,L\}$ our expressions fully agree with those given in \cite{Laenen:1992zk} except for
a wrong sign in front of the $\ln(\chi)^2$ in their Eq.~(3.25) (previously also found by \cite{Harris:1995tu}).
Collecting all prefactors, the double differential partonic PGF cross section with one additional
soft gluon emission is given by
\begin{align}
\nonumber
&{s'}^2\frac{d^2\sigma_{k,\Pg}^{(1),S}}{dt_1du_1} 
= \alpha\alpha_s^2 K_{\Pg\Pgg} b_k(\epsilon)
\frac{2^8\pi^4 S_\epsilon}{\Gamma(1+\epsilon/2)} \left(\frac{\mu_D^2}{m^2}\right)^{-\epsilon} \\
\nonumber
& \left(\frac{(t_1u_1'-s'm^2)s' - q^2t_1^2}{m^2{s'}^2}\right)^{\epsilon/2} 
C_\epsilon  \left(\frac \Delta {m^2}\right)^\epsilon \delta(s'+t_1+u_1)\\
\nonumber
&N_C\, C_F\, E_k(\epsilon) \, e_H^2\,
\frac{s_4}{2\pi(s_4+m^2)}\frac{s_4}{\epsilon}\left[1-\frac 3 8 \zeta(2)\epsilon^2\right]\\
&\int\!d\Omega_n \,d\hat{\mathcal I}\,\left(C_A S_{k,\tOK} + 2C_F S_{k,\tQED}\right)
\label{eq:soft-xsec}
\end{align}
with the expressions for $S_{k,\tOK}$ and $S_{k,\tQED}$ 
given in Eqs.~(\ref{eq:soft-qed}) and (\ref{eq:soft-ok}), respectively, to be inserted in the last line.
Upon adding the renormalized virtual cross section to Eq.~(\ref{eq:soft-xsec}) all
$1/\epsilon^2$ and $1/\epsilon$ singularities of IR origin cancel, and the
remaining collinear divergency in the OK part will be removed by mass factorization
to be discussed in Sec.~\ref{sec:massfact} below.

As was mentioned above, our partonic cross sections have been derived for an observed
heavy antiquark $\PaQ$. Since the PGF cross section is symmetric under the
exchange of $p_1$ and $p_2$, the results do not change if 
the heavy quark $\PQ$ is observed instead.
\subsection{Light Quark Initiated Processes \label{sec:light-quarks}}
At NLO accuracy, also light-quark initiated processes, 
\begin{equation}
\label{eq:proc-quarks}
\Pggx(q) + \Pq(k_1) \rightarrow \PQ(p_1)+\PaQ(p_2) + \Pq(k_2) \;,
\end{equation}
have to be considered. The two contributing mechanisms are depicted in 
Fig.~\ref{fig:feyn-quark} and differ by the electrical charge coupling
to the virtual photon. 
\begin{figure}[ht!]
\begin{center}
\includegraphics[width=0.45\textwidth]{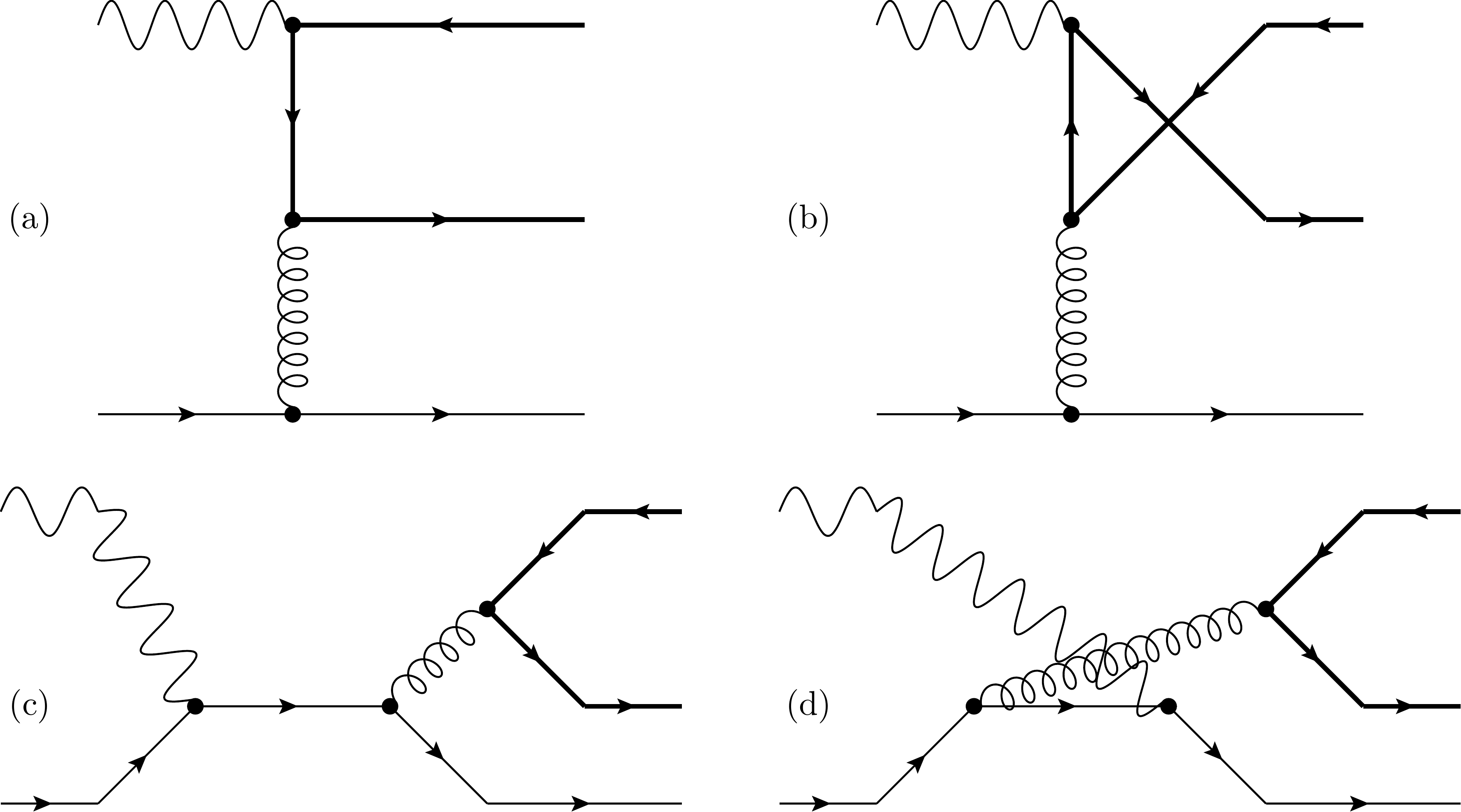}
\end{center}
\vspace*{-0.3cm}
\caption{Feynman diagrams (a), (b) and (c), (d) denote the 
light quark initiated Bethe-Heitler and Compton process, respectively,
that start to contribute to HQ electroproduction at NLO accuracy.
Similar contributions arise for incoming light antiquarks.}\label{fig:feyn-quark}
\end{figure}
The Bethe-Heitler process is proportional to
the charge $e_H$ of the produced HQ whereas the charge $e_L$
of the incoming light-quark is relevant for Compton scattering.
The NLO matrix element squared for (\ref{eq:proc-quarks}), 
properly projected onto the polarization $k$, Lorentz-, and Dirac-structure, 
can be, hence, decomposed into three pieces $A_{k,i}$, $i=1,2,3$, according to the 
electrical charges $e_H$ and $e_L$ (in units of $e$):
\begin{eqnarray}
\nonumber
\hat {\mathcal P}_{k}^{\Pgg,\mu\mu'}
\hat {\mathcal P}_{k}^{\Pq,aa'} &&\!\!\!\!\!\!\!\!
\sum_{j,j'=1}^4\Md^{(1),q}_{j,\mu a}\left(\Md^{(1),q}_{j',\mu' a'}\right)^* =
8g^4\mu_D^{-2\epsilon}e^2 N_C C_F \\ 
&\times&\left( e_H^2\, A_{k,1} +  e_L^2\, A_{k,2} +  e_L e_H\, A_{k,3} \right)\;.
\label{eq:quark-aki}
\end{eqnarray}

At this order, one encounters only collinear divergencies, which can be solely attributed
to the Bethe-Heitler process $A_{k,1}$. With the $n$-dimensional $2\to 3$ phase space
given in the previous subsection, the singularity structure of $A_{k,1}$ reads
\begin{eqnarray}
\nonumber
&&\lefteqn{\frac{s_4}{2\pi(s_4+m^2)}\int\!d\Omega_n d\hat{\mathcal I}\,C_F A_{k,1} =} \\
&&-\frac 1 {u_1}B_{k,\tQED}\left(\!\begin{array}{l}s'\rightarrow x_1s'\\t_1\rightarrow x_1 t_1\end{array}\!\!\right) 
P^{k,(0)}_{\Pg\Pq}(x_1)\frac 2 \epsilon + {\mathcal O}(\epsilon^0)
\label{eq:quark-coll}
\end{eqnarray}
with, as before, $x_1 = -u_1/(s'+t_1)$ and, depending on the projection $k$, the 
appropriate LO splitting kernels  \cite{Altarelli:1977zs}
\begin{align}
P^{G,(0)}_{\Pg\Pq}(x) = P^{L,(0)}_{\Pg\Pq}(x) &= C_F\left(\frac 1 x + \frac {(1-x)^2} x\right)\;, \label{eq:PGLgq} \\
P^{P,(0)}_{\Pg\Pq}(x) &= C_F\left(2-x\right)\;. \label{eq:PPgq}
\end{align}
Again, the $1/\epsilon$ pole will be dealt with by mass factorization 
as we shall discuss in Section \ref{sec:massfact} below.
$A_{k,2}$ does not develop a collinear singularity as long as the probing photon is virtual, i.e.,
$q^2\neq 0$. In the limit of photoproduction, see Refs.~\cite{Smith:1991pw,Bojak:1998bd}, an additional
mass factorization is required to arrive at a finite expression for the process (\ref{eq:proc-quarks}).

The $n$-dimensional partonic cross section for (\ref{eq:proc-quarks}) at NLO accuracy, 
differential in $t_1$ and $u_1$, is given by
\begin{align}
&{s'}^2\frac{d^2\sigma_{k,\Pq}^{(1)}}{dt_1du_1} = \alpha\alpha_s^2 K_{\Pq\Pgg} b_k(\epsilon)
\frac{2^7 \pi^3 S_\epsilon^2}{\Gamma(1+\epsilon)} \left(\frac{\mu_D^2}{m^2}\right)^{-\epsilon} \frac{s_4}{s_4+m^2}
 \nonumber\\
&  \left(\frac{(t_1u_1'-s'm^2)s' - q^2t_1^2}{m^2{s'}^2}\right)^{\epsilon/2} 
\left(\frac{s_4^2}{m^2(s_4+m^2)}\right)^{\epsilon/2}  \nonumber\\
&N_C\, C_F
\int\!d\Omega_n d\hat{\mathcal I}\,\left(e_H^2 A_{k,1} + e_L^2 A_{k,2} + e_H e_L A_{k,3} \right)
\label{eq:xsec-lightquark}
\end{align}
with the color average $K_{\Pq\Pgg} = 1/N_C$.
We note that the finite interference contribution $A_{k,3}$ will only contribute to HQ electroproduction
as long as one does not fully integrate over phase space, which is a consequence of Furry's theorem,
i.e., $\int\!dt_1du_1\int\!d\Omega_n d\hat{\mathcal I}\,A_{k,3}=0$.

As was also stressed in case of photoproduction \cite{Smith:1991pw,Bojak:1998bd},
the Mandelstam variables $t_1$ and $u_1$ are defined for the momentum transfer of an
observed heavy \textit{anti}quark $\bar{Q}(p_2)$ with respect to the incoming virtual photon and light quark.
Detecting the heavy quark $Q(p_1)$ instead amounts to interchanging $t_1 \leftrightarrow u_1$. Since
both $A_{k,1}$ and $A_{k,2}$ are symmetric in $t_1$ and $u_1$ nothing changes. 
However, one needs to take into account an overall change of sign for the interference
term $A_{k,3}$ that is purely antisymmetric under $t_1 \leftrightarrow u_1$.
Replacing the incoming light quark in (\ref{eq:proc-quarks}) by a light antiquark
is taken into account by the following identities \cite{Smith:1991pw,Bojak:1998bd}
\begin{eqnarray}
{s'}^2\frac{d^2\sigma_{k,\Pq}^{(1)}}{dt_1du_1}\Big|_{\Pggx\Paq\to \PaQ} &=&
{s'}^2\frac{d^2\sigma_{k,\Pq}^{(1)}}{dt_1du_1}\Big|_{\Pggx\Pq\to \PQ}\;,\\
{s'}^2\frac{d^2\sigma_{k,\Pq}^{(1)}}{dt_1du_1}\Big|_{\Pggx\Paq\to \PQ} &=&
{s'}^2\frac{d^2\sigma_{k,\Pq}^{(1)}}{dt_1du_1}\Big|_{\Pggx\Pq\to \PaQ}\;.
\end{eqnarray}
\subsection{Mass Factorization Procedure \label{sec:massfact}}
In the last step, one needs to remove the remaining collinear singularities in
the PGF and the Bethe-Heitler processes that we have encountered in Sections
\ref{sec:realcorr} and \ref{sec:light-quarks} in Eqs.~(\ref{eq:gluon-coll}) and
(\ref{eq:quark-coll}), respectively.
To this end, a standard mass factorization procedure at NLO needs to be applied,
which absorbs the remaining collinear divergencies into the definition of the parton
distribution functions at a factorization scale $\mu_F$, yielding finite
(reduced) partonic cross sections, where the limit $\epsilon \to 0$ can be taken.

In case of the PGF process, mass factorization amounts to 
\begin{align}
\nonumber
&\lefteqn{{s'}^2 \frac{d^2\sigma_{k,\Pg}^{(1)}(s',t_1,u_1,\mu_F)}{dt_1du_1} =} \\ 
\nonumber
&\lim_{\epsilon\rightarrow 0}\Bigg[ {s'}^2\frac{d^2\sigma_{k,\Pg}^{(1)}(s',t_1,u_1,\epsilon)}{dt_1du_1} 
-\int\limits_0^1\frac{dx_1'}{x_1'}\Gamma_{\Pg\Pg}^{k,(1)}(x_1',\mu_F,\mu_D,\epsilon)\\
&\times (x_1's')^2 \; \frac{d^2\sigma_{k,\Pg}^{(0)}(x_1's',x_1't_1,u_1,\epsilon)}{d(x_1t_1)du_1} \Bigg]\;,
\label{eq:gluon-fact}
\end{align}
i.e., subtracting the convolution of the gluon-gluon transition function at NLO
\begin{align}
\label{eq:trans-gg}
\Gamma_{\Pg\Pg}^{k,(1)}(x,\mu_F,\mu_D,\epsilon) = \frac{\alpha_s}{2\pi}\left[P^{k,(0)}_{\Pg\Pg}(x)\frac{2}{\epsilon} 
+ f^{k,(1)}_{\Pg\Pg}(x,\mu_F^2,\mu_D^2)\right]
\end{align}
and the $n$-dimensional Born cross section $d^2\sigma_{k,\Pg}^{(0)}$ in Eq.~(\ref{eq:born-ndim})
at an appropriately rescaled kinematics $x_1 s'$ and $x_1 t_1$ off the collinear singular NLO cross section;
to avoid any confusion with the shifted collinear kinematics, we keep in this section the relevant arguments in all 
expressions for partonic cross sections whenever necessary, such as in Eq.~(\ref{eq:gluon-fact}).
In Eq.~(\ref{eq:trans-gg}), the finite function $f^{k,(1)}_{\Pg\Pg}$ carries the choice of factorization scheme,
i.e., the finite term that will be subtracted along with the $1/\epsilon$ pole,
and the dependence on the factorization scale $\mu_F$. In the conventional
$\overline{\text{MS}}$ scheme, which we adopt, it reads
\begin{align}
\nonumber
f^{k,(1)}_{\Pg\Pg}(x,\mu_F^2,\mu_D^2) &= P^{k,(0)}_{\Pg\Pg}(x) \Bigg[
\gamma_E - \ln(4\pi) + \ln\left(\frac{\mu_F^2}{m^2}\right) \\
&- \ln\left(\frac{\mu_D^2}{m^2}\right)\Bigg]
\end{align}
with the LO gluon-to-gluon splitting function
\begin{align}
\nonumber
P^{k,(0)}_{\Pg\Pg}(x) &= \Theta(1-\delta-x) P^{k,(0)}_{\Pg\Pg,H}(x) \\
&+\delta(1-x)\left(2\,C_A\ln(\delta)+\frac{\beta_0^{lf}}{2}\right)
\label{eq:pgg0}
\end{align}
where $\beta_0^{lf} = (11C_A - 2n_{lf})/3$.
Here, we have introduced another infrared cut-off $\delta$ to separate 
soft ($x\geq 1-\delta$) and hard ($x<1-\delta$) collinear gluons. $\Delta$ is related
to $\delta$ by simple kinematics through the relation $\delta=\Delta/(s'+t_1)$;
see, also Ref.~\cite{Laenen:1992zk}.  The hard part
$P^{k,(0)}_{\Pg\Pg,H}(x)$ was already given in Eq.~(\ref{eq:pgg-hard})
to specify the contribution of collinear pole to $R_{k,\tOK}$ in Eq.~(\ref{eq:gluon-coll}).
$\delta$ is used to split the finite reduced PGF cross section again 
into a hard and a virtual plus soft contribution. 
 
In general, the $\overline{\text{MS}}$ transition functions at NLO accuracy take the following form 
for each projection $k$
\begin{align}
\nonumber
&\Gamma_{ij}^{k,(1)}(x,\mu_F,\mu_D,\epsilon) = \frac{\alpha_s}{2\pi} P^{k,(0)}_{ij}(x)
\Bigg[\frac{2}{\epsilon} +\gamma_E - \ln(4\pi)\\ 
\nonumber
& + \ln\left(\frac{\mu_F^2}{m^2}\right) - \ln\left(\frac{\mu_D^2}{m^2}\right)\Bigg]\\
&= 8\pi\alpha_s\, P^{k,(0)}_{ij}(x)\, C_\epsilon 
\left(\frac{\mu_D^2}{m^2}\right)^{-\epsilon/2} \left[\frac{2}{\epsilon} + 
\ln\left(\frac{\mu_F^2}{m^2}\right)\right]\;,
\end{align}
which we use to regularize the collinear singularities in the light-quark initiated
Bethe-Heitler process, see Eq.~(\ref{eq:quark-coll}),
\begin{align}
\nonumber
&\lefteqn{{s'}^2\frac{d^2\sigma_{k,\Pq}^{(1)}(s',t_1,u_1,\mu_F)}{dt_1du_1} =}\\
\nonumber 
&\lim_{\epsilon\rightarrow 0} \Bigg[{s'}^2\frac{d^2\sigma_{k,\Pq}^{(1)}(s',t_1,u_1,\epsilon)}{dt_1du_1} 
  -\int\limits_0^1\frac{dx_1}{x_1} \Gamma^{k,(1)}_{\Pg\Pq}(x_1,\mu_F,\mu_D,\epsilon)\\
&\times (x_1s')^2 \, \frac{d^2\sigma_{k,\Pg}^{(0)}(x_1s',x_1t_1,u_1,\epsilon)}{d(x_1t_1)du_1} \Bigg]\;. 
\label{eq:quark-fact}
\end{align}
Here, the quark-to-gluon $P^{k,(0)}_{\Pg\Pq}$ splitting function, see Eqs.~(\ref{eq:PGLgq}) and (\ref{eq:PPgq}),
is needed.

We can now quote the final, finite expressions for the gluon and light-quark initiated processes contributing 
to HQ production in DIS at NLO accuracy for all spin projections $k\in \{G,L,P\}$. 
The PGF result, split into hard and soft plus virtual contributions, reads
\begin{widetext}
\begin{align}
\nonumber
{s'}^2\frac{d^2\sigma_{k,\Pg}^{(1),H}}{dt_1du_1} 
&=\frac{1}{2\pi}\,K_{\Pg\Pgg}\,\alpha\,\alpha_S \,e_H^2 \, N_C\,C_F\,b_k(0)\, \Bigg\{-\frac 1 {u_1}P^{k,(0)}_{\Pg\Pg,H}(x_1) 
\Bigg[ 4\pi\, B^{(0)}_{k,\tQED} \left(\!\begin{array}{l}s'\rightarrow x_1s'\\t_1\rightarrow x_1 t_1\end{array}\!\!\right) \\
\nonumber
&\times \left[ \ln\left(\frac{s_4^2}{m^2(s_4+m^2)}\right) - \ln\left(\frac{\mu_F^2}{m^2}\right)  \right]
-8\pi\, B^{(1)}_{k,\tQED}\left(\!\begin{array}{l}s'\rightarrow x_1s'\\t_1\rightarrow x_1 t_1\end{array}\!\!\right)\Bigg] \\
&+C_A\frac{s_4}{s_4+m^2}\left(\int\!d\Omega_n \, d\hat{\mathcal I}\,R_{k,\tOK}\right)^{\text{finite}}
+2\,C_F\frac{s_4}{s_4+m^2}\int\!d\Omega_4 \, d\hat{\mathcal I}\,R_{k,\tQED}\Bigg\}
\label{eq:pgf-hard}
\end{align}
and
\begin{align}
\nonumber
{s'}^2\frac{d^2\sigma_{k,\Pg}^{(1),S+V}}{dt_1du_1} &= 4\,K_{\Pg\Pgg}\, \alpha\,\alpha_S \, e_H^2\, N_C\,C_F\,b_k(0)\,
B^{(0)}_{k,\tQED}\, \delta(s'+t_1+u_1) \Bigg\{ C_A \ln^2\left(\frac{\Delta}{m^2}\right) + \ln\left(\frac{\Delta}{m^2}\right)
\\
\nonumber
&\times \left[ C_A \left[\ln\left(\frac{-t_1}{m^2}\right)-\ln\left(\frac{-u_1}{m^2}\right)-
\ln\left(\frac{\mu_F^2}{m^2}\right)\right]  
- (C_A-2C_F)  \frac{2m^2-s}{s\beta} \ln(\chi)- 2\,C_F\right] \nonumber\\
&+ C_A \ln\left(\frac{\mu_F^2}{m^2}\right)\ln\left(\frac{-u_1}{m^2}\right) + \frac {\beta_0^{lf}}{4}\left[ \ln\left(\frac{\mu_R^2}{m^2}\right)- \ln\left(\frac{\mu_F^2}{m^2}\right)\right] +  f_k(s',u_1,t_1,m^2,q^2)\Bigg\}\;,
\label{eq:pgf-s+v}
\end{align}
respectively. Here, the $B^{(l)}_{k,\tQED}$ denote the $\mathcal O(\epsilon^l)$ coefficients to 
the $n$-dimensional Born cross section
$B_{k,\tQED}$ given in  Eqs.~(\ref{eq:BQEDG}) - (\ref{eq:BQEDP}), i.e., we write
\begin{equation}
B_{k,\tQED} = B^{(0)}_{k,\tQED} + \epsilon\, B^{(1)}_{k,\tQED} + \epsilon^2\, B^{(2)}_{k,\tQED}\;.
\end{equation}
Recall that for $k=\{L,P\}$ there are no contributions proportional to $\epsilon$, i.e., $B^{(1)}_{L,\tQED} = B^{(1)}_{P,\tQED} = 0$.
As was mentioned above, the QED part, $R_{k,\tQED}$ in (\ref{eq:pgf-hard}), 
does not require any mass factorization at this order.
The functions $f_k$ in (\ref{eq:pgf-s+v}) contain logarithms and dilogarithms with different, complicated
arguments, but they do not depend on $\Delta$, $\mu_F^2$, $\mu_R^2$ nor on $n_f$ and $\beta_0^{lf}$.

The corresponding finite, reduced partonic cross section for the light quark initiated Bethe-Heitler and
Compton processes reads
\begin{align}
\nonumber
{s'}^2\frac{d^2\sigma_{k,\Pq}^{(1)}}{dt_1du_1} &=\frac{1}{2\pi}\,K_{\Pq\Pgg}\,\alpha\,\alpha_S\,N_C\,b_k(0)
\Bigg\{ -\frac 1 {u_1}e_H^2 P^{k,(0)}_{\Pg\Pq}(x_1) \Bigg[ 
2\pi\, B^{(0)}_{k,\tQED} \left(\!\begin{array}{l}s'\rightarrow x_1s'\\t_1\rightarrow x_1 t_1\end{array}\!\!\right) \\
\nonumber
&\times \left[ \ln\left(\frac{s_4^2}{m^2(s_4+m^2)}\right)-\ln(\mu_F^2/m^2)+1-\delta_{k,P}\right]
-4\pi\, B^{(1)}_{k,\tQED} \left(\!\begin{array}{l}s'\rightarrow x_1s'\\t_1\rightarrow x_1 t_1\end{array}\!\!\right) \Bigg] \\
&+C_F\frac{s_4}{s_4+m^2} \left[
\left(\int\!d\Omega_nd\hat{\mathcal I}\,e_H^2A_{k,1}\right)^{\text{finite}}
+\int\!d\Omega_4d\hat{\mathcal I}\,e_L^2A_{k,2} 
+\int\!d\Omega_4d \hat{\mathcal I}\,e_H e_L A_{k,3} \right] \Bigg\}\;,
\label{eq:quark-finite}
\end{align}
\end{widetext}
where $1-\delta_{k,P}$ may also be written as $-2\partial_\epsilon E_{k}(\epsilon=0)$ 
as it originates from the additional factor of $E_k(\epsilon)$ hidden in the 
subtraction piece in Eq.~(\ref{eq:quark-fact}). Again, recall that
neither $A_{k,2}$ nor $A_{k,3}$ require any mass factorization at NLO as long as the photon is not on mass
shell.

The finite expressions corresponding to $R_{k,\tOK}$ and $R_{k,\tQED}$ in Eq.~(\ref{eq:pgf-hard})
and for $f_k$ in Eq.~(\ref{eq:pgf-s+v}), differential in $t_1$ and $u_1$,
are too lengthy to be reproduced here, but they are available upon request; similarly, for
the light-quark contributions $A_{k,1}$, $A_{k,2}$, and $A_{k,3}$ in Eq.~(\ref{eq:quark-finite}).
We note that for unpolarized DIS, $k=\{G,L\}$, we fully agree with the expressions given in Ref.~\cite{Laenen:1992zk}.
The complete NLO result for longitudinal polarization, $k=P$, is new and for the first
time given in this paper. We shall comment on further extensive 
comparisons to existing results, both analytically and numerically, in the next section.
\section{Total Partonic Cross Sections \label{sec:partonic}}
In the previous section we have obtained all the ingredients to compute
the double-differential partonic cross sections $d^2\sigma^{(n)}_{k,j}/(dt_1 du_1)$
for HQ electroproduction at NLO accuracy for all projections $k=\{G,L,P\}$.
Upon convolution with appropriate combinations of PDFs this will yield, for instance, 
results for transverse momentum and rapidity distributions of an observed heavy
antiquark (or quark) in DIS at a given $x$ and $Q^2$. 
We will pursue this type of DIS observables further 
in a forthcoming publication \cite{ref:inprep}.

In this paper, we are mainly interested in the impact of the NLO corrections for $k=P$
on the longitudinally polarized inclusive DIS structure function $g_1^{\PQ}$, 
and the corresponding, experimentally relevant double-spin asymmetry 
commonly defined as
\begin{equation}
\label{eq:spin-asym}
A_1^{\PQ}(x,Q^2,m^2) = \frac{g_1^{\PQ}(x,Q^2,m^2)}{F_1^{\PQ}(x,Q^2,m^2)}\;, 
\end{equation}
with the unpolarized structure function $F_1^{\PQ}$ in the denominator.
The ratio (\ref{eq:spin-asym}) has the virtue that
some sources of experimental uncertainties are conveniently expected to drop out.
Therefore, we proceed by computing the related \textit{total} partonic cross sections 
$\sigma_{k,j}(s,q^2,m^2)$ up to NLO accuracy, which are obtained by integrating 
the differential expressions derived in the previous section
over the entire kinematic range for fixed $s$ and $q^2$.
To this end, it is convenient to trade, for instance, $u_1$ for the 
Mandelstam variable $s_4= s'+t_1+u_1$ which controls the soft limit. One obtains
\begin{align}
\nonumber
\sigma^{(n)}_{k,j}(s,q^2,m^2) &= 
\!\!\!\!\! \int\limits_{-s'(1+\beta)/2}^{-s'(1-\beta)/2} 
\!\!\!\! dt_1 \!\! 
\int\limits_0^{s_{4,\max}} \!\!ds_4 \frac{d^2\sigma^{(n)}_{k,j}(s',t_1,u_1,q^2)}{dt_1ds_4}\\
\label{eq:totalpartonic}
\end{align}
where the upper limit of $s_4$ is given by
\begin{equation}
s_{4,\max} = \frac{s}{s' t_1}\left(t_1+\frac{s'(1-\beta)}{2}\right)
            \left(t_1+\frac{s'(1+\beta)}{2}\right)\;.
\end{equation}
The total partonic cross section at NLO accuracy is then obtained by adding the Born ($n=0$) result
and the $\mathcal{O}(\alpha_s)$ corrections $(n=1)$, i.e.,
\begin{equation}
\label{eq:tot-nlo}
\sigma_{k,j}(s,q^2,m^2) = \sigma^{(0)}_{k,j}(s,q^2,m^2) + \sigma^{(1)}_{k,j}(s,q^2,m^2)\;.
\end{equation}
As before, $k$ denotes the projection $G$, $L$, and $P$ onto the relevant unpolarized
and longitudinally polarized HQ cross sections, respectively, and $j\in\{\Pg,\Pq,\Paq\}$ 
labels the flavor of the incoming parton.
Upon convolution of (\ref{eq:tot-nlo}) with PDFs one obtains the DIS HQ structure functions
at NLO accuracy, in particular, $g_1^{\PQ}$ for $k=P$, as we shall discuss in detail in
Sec.~\ref{sec:pheno}.

At LO accuracy, the $dt_1 du_1$ or $dt_1 ds_4$ integrations in Eq.~(\ref{eq:totalpartonic})
are straightforwardly performed, see also Eq.~(\ref{eq:born-ndim}), 
and one obtains for $k=\{G,L\}$ and $k=P$
\begin{align}
\sigma_{L,\Pg}^{(0)}(s,q^2,m^2) &= 16\pi\alpha\alpha_s e_H^2 K_{\Pg\Pgg}N_CC_F
\left(\frac{-q^2s}{{s'}^3}\right) \nonumber \\
&\times \left[\beta + \frac{2m^2}{s}\ln(\chi)\right]\;,\\
\sigma_{G,\Pg}^{(0)}(s,q^2,m^2) &=
-4\pi\alpha\alpha_s e_H^2 K_{\Pg\Pgg}N_CC_F\frac 1 {{s'}^3} \nonumber \\
&\times \Big[(s^2+q^4+4m^2 s)\beta \nonumber \\ 
&+ (s^2+q^4-4m^2(2m^2-s'))\ln(\chi)\Big]\;, \\
\sigma_{T,\Pg}^{(0)}(s,q^2,m^2) &= \sigma_{G,\Pg}^{(0)}(s,q^2,m^2) + \frac{1}{2} \sigma_{L,\Pg}^{(0)}(s,q^2,m^2)\;, \\
\sigma_{P,\Pg}^{(0)}(s,q^2,m^2) &= 4\pi\alpha\alpha_s e_H^2 K_{\Pg\Pgg}N_CC_F \nonumber \\
&\times \frac 1 {{s'}^2} \left[(3s+q^2)\beta + (s+q^2)\ln(\chi)\right]\;,
\end{align}
in agreement with Ref.~\cite{Laenen:1992zk} and \cite{Watson:1981ce,Vogelsang:1990ug}, respectively. 
As in Eq.~(\ref{eq:def-trans}), the transverse partonic cross section is obtained from
the expressions for $k=G$ and $k=L$.

At NLO accuracy, it is customary \cite{Laenen:1992zk} to further decompose
Eq.~(\ref{eq:tot-nlo}) as follows
\begin{eqnarray}
\nonumber
\lefteqn{\sigma_{k,j}(s,q^2,m^2) = \frac{\alpha\alpha_s}{m^2}  \left[ f_{k,j}^{(0)}(\eta,\xi) 
 + 4\pi\alpha_s \Big(f_{k,j}^{(1)}(\eta,\xi) \right. }\\
&+& \left.  \ln(\mu_F^2/m^2)\bar f_{k,j}^{F,(1)}(\eta,\xi) + \ln(\mu_R^2/m^2)\bar f_{k,j}^{R,(1)}(\eta,\xi)\Big)\right]
\nonumber \\
\label{eq:partonic-xsec}
\end{eqnarray}
where each function $f^{(n)}_{k,j}$ only depends on the scaling variables $\eta=1/\rho-1$ and $\xi=-q^2/m^2$. 
In addition, any global dependence on the electrical charges $e_H$ and $e_L$ of the heavy and light quarks
scattering off the virtual photon, respectively, is usually factored out, yielding 
\begin{align}
\label{eq:partonic-fsplit-gluon}
f_{k,\Pg}^{(n)}(\eta,\xi) &= e_H^2 c_{k,\Pg}^{(n)}(\eta,\xi) \;,\\
\label{eq:partonic-fsplit-quark}
f_{k,\Pq}(\eta,\xi) &= e_H^2 c_{k,\Pq}(\eta,\xi) + e_L^2 d_{k,\Pq}(\eta,\xi)\;,
\end{align}
with similar expressions for the functions $\bar f_{k,j}^{F,(1)}$ and $\bar f_{k,j}^{R,(1)}$
in Eq.~(\ref{eq:partonic-xsec})
in terms of $\bar c_{k,j}^{F,(1)}$ and $\bar d_{k,\Pq}^{F,(1)}$. The latter
multiply the logarithmic dependence on the factorization and renormalization scales, respectively,
if $\mu_F$ and/or $\mu_R$ are chosen different from the HQ mass $m$. 
Due to Furry's theorem, the quark coefficient proportional to $e_H e_L$, present in differential
cross sections, see Eq.~(\ref{eq:quark-aki}), vanishes when integrated over the entire phase space.

The renormalization scale dependence trivially arises from the renormalization of
the strong coupling. The corresponding coefficient function $\bar c_{k,\Pg}^{R,(1)}$ at NLO accuracy
is proportional to the Born coefficient $c_{k,\Pg}^{(0)}$ and the QCD beta function evaluated with
$n_{lf}$ light flavors, i.e., one finds
\begin{equation}
\bar c_{k,\Pg}^{R,(1)} = \frac{\beta_0^{lf}}{16\pi^2} c_{k,\Pg}^{(0)}\;.
\label{eq:cgBarR}
\end{equation}
In what follows, we always present results for the sum
\begin{equation}
\label{eq:bar-fr-sum}
\bar c_{k,\Pg}^{(1)} = \bar c_{k,\Pg}^{F,(1)} + \bar c_{k,\Pg}^{R,(1)}
\end{equation}
instead of the individual pieces. This yields the
shortest expressions for the usual choice of common factorization and renormalization
scales, i.e., $\mu_F=\mu_R$, since in this case the 
dependence on the QCD beta function $\beta_0^{lf}$ cancels in the sum (\ref{eq:bar-fr-sum}).
Since the quark coefficients are genuine NLO corrections, they do not
carry any renormalization dependence at this order in pQCD, i.e.,
$\bar{c}_{k,\Pq}^{R,(1)}=\bar{d}_{k,\Pq}^{R,(1)}=0$.
In addition, $d_{k,\Pq}^{(1)}$ is also
free of collinear singularities, yielding $\bar{d}_{k,\Pq}^{F,(1)}=0$,
as long as $Q^2\neq 0$, i.e., away from the limit of photoproduction.
In App.~\ref{app:scaling-fcts}, we give, whenever possible, compact analytic results for the 
partonic scaling functions. Otherwise, results 
are available upon request from the authors.

At this point, a brief digression about the numerical implementation of the 
phase space slicing method is in order.  The technique was adopted in the analytical 
calculations to split the PGF cross section into contributions from
hard ($s_4>\Delta$) and soft ($s_4<\Delta$) gluon radiation, with the latter being
added to the virtual contributions to cancel all IR singularities analytically.
One needs to ensure that $\Delta$ is sufficiently small with respect to the
$2\to 2$ Mandelstam variables and the HQ mass $m^2$ to be negligible. In practice,
see also Refs.~\cite{Smith:1991pw,Laenen:1992zk,Bojak:1998bd}, 
$\Delta \simeq \left(10^{-5}\ldots 10^{-7}\right) m^2$, 
guarantees numerically stable results, i.e., the cancellation of logarithms in
$\Delta/m^2$. This is further enforced by rewriting the soft plus virtual
cross section, expanded in powers of $\ln^i (\Delta/m^2)$, $i=0,\,1$, and 2,
with the help of the identity \cite{Beenakker:1990maa,Bojak:1998bd}
\begin{equation}
\delta(s_4) \sum_{i=0}^2 \alpha_i \ln^i \left(\frac{\Delta}{m^2}\right) = \Theta(s_4-\Delta)\sum_{i=0}^2
{\cal{A}}_i \;\alpha_i|_{s4=0}
\end{equation}
with expansion coefficients
\begin{eqnarray}
{\cal{A}}_0 &=& \frac{1}{s_{4,\max}-\Delta} \;,\nonumber\\
{\cal{A}}_1 &=& \frac{\ln(s_{4,\max}/m^2)}{s_{4,\max}-\Delta} - \frac{1}{s_4} \;,\nonumber\\
{\cal{A}}_2 &=& \frac{\ln^2 (s_{4,\max}/m^2)}{s_{4,\max}-\Delta} - \frac{2\ln(s_4/m^2)}{s_4} \;,
\end{eqnarray}
before being added to the hard gluon part.

To ensure the correctness of our new results for longitudinally polarized DIS, i.e., for $k=P$, 
relevant for the computation of the HQ structure function $g_1^{\PQ}$ at NLO accuracy, 
we have performed extensive comparisons, both
analytically and numerically, to the mainly unpolarized results for the HQ scaling functions
already available in the literature \cite{Laenen:1992zk,Riemersma:1994hv,Harris:1995tu}. 
These checks include results for 
the threshold \cite{Laenen:1992zk,Laenen:1998kp,Eynck:2000gz,Kawamura:2012cr}
and high-energy \cite{Catani:1990eg} limit, $\eta\to 0$ and $\eta\to\infty$, respectively, 
and the limit $Q^2\gg m^2$ \cite{Buza:1995ie,Buza:1996xr}
that is relevant for the construction of GM-VFNS \cite{Buza:1996wv} in PDF analyses. 
Both, in the unpolarized and the polarized case, we can also compare our results to the known 
limit of photoproduction \cite{Smith:1991pw,Bojak:1998bd} 
except for the light-quark scaling function $d^{(1)}_{k,\Pq}$, which is singular for $\xi\to 0$
and would require an additional mass factorization into a contribution from the hadronic structure
of real photons for $\xi=0$. 
Unless stated otherwise below, we fully agree with the literature.
We note that our results for the Compton process, $d^{(1)}_{k,\Pq}(\eta,\xi)$, 
see App.~\ref{app:scaling-fcts}, match for all 
projections $k$ analytically with the corresponding expressions 
recently derived in Ref.~\cite{Blumlein:2016xcy}.

\begin{figure}[ht!]
\begin{center}
\includegraphics[width=0.48\textwidth]{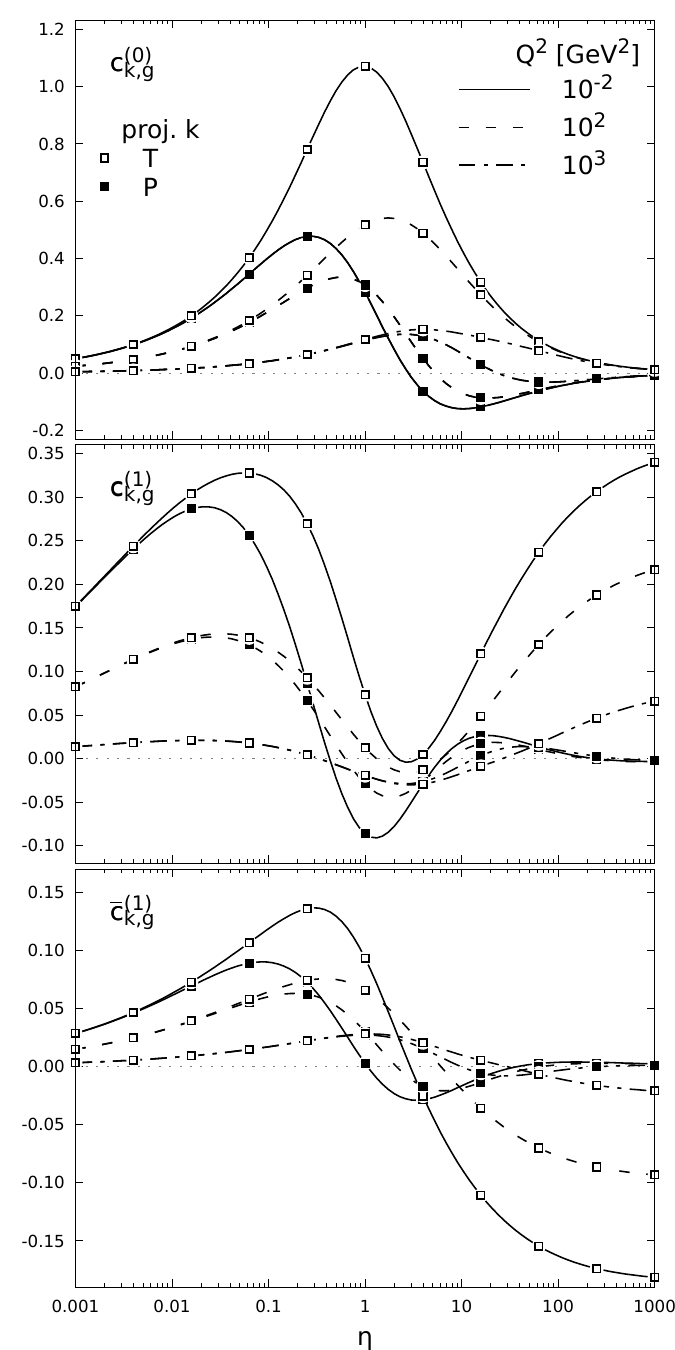}
\end{center}
\vspace*{-0.5cm}
\caption{\label{fig:scaling-gluon} The top, middle, and lower panel show the 
Born $c_{k,\Pg}^{(0)}$, and the NLO $c_{k,\Pg}^{(1)}$ and $\bar{c}_{k,\Pg}^{(1)}$ 
contributions, respectively, to the total cross section $\sigma_{k,\Pg}$ at NLO
in Eq.~(\ref{eq:partonic-xsec}) 
as a function of $\eta$ for three different values of $Q^2$ and fixed 
HQ mass $m=\SI{4.75}{\GeV}$. 
The open and closed symbols denote the projections $k=T$ and $k=P$, respectively.}
\end{figure}
Figures \ref{fig:scaling-gluon} and \ref{fig:scaling-quark} show the polarized ($k=P$) 
gluonic and light-quark scaling functions defined in Eqs.~(\ref{eq:partonic-fsplit-gluon}) 
and (\ref{eq:partonic-fsplit-quark}), respectively, relevant for the calculation
of $g_1^{\PQ}$ in Sec.~\ref{sec:pheno}.
Results for $k=T$, needed for computing the denominator $F_1^{\PQ}$ 
in the double-spin asymmetry $A_1^{\PQ}$, are also given for comparison for the dominant PGF process.
In case of the light-quark scaling functions shown in Fig.~\ref{fig:scaling-quark},
the unpolarized results for $k=T$ turn out to be numerically
significantly larger than those for $k=P$. Hence, it is impossible to
display them together. Here, and in general for the scaling functions for $k=L$, 
we refer the reader to Ref.~\cite{Laenen:1992zk} for the corresponding plots.    
We note, that we fully agree numerically with all the results
for the unpolarized scaling functions given in Ref.~\cite{Laenen:1992zk} 
except for some of the curves for $d^{(1)}_{L,\Pq}$ shown in their Fig.~11~(b) 
which seem to be mislabeled (this error has also been found in Ref.~\cite{Harris:1995tu}).
As in Ref.~\cite{Laenen:1992zk}, our results in 
Figs.~\ref{fig:scaling-gluon} and \ref{fig:scaling-quark} are shown as function of $\eta$
for a fixed HQ mass of $m=\SI{4.75}{\GeV}$ and for several representative
values of $Q^2$, which approximately span the range of 
$0.44\times10^{-3}\le \xi=Q^2/m^2\le 44$.

In general, all scaling functions exhibit a rather non-trivial dependence on $\eta$ that has been
discussed at some length in the unpolarized case in Ref.~\cite{Laenen:1992zk}. Let us only
add few observations concerning the similarities and differences in the behavior of
the polarized results derived for the first time in this paper. 
Most importantly, one finds for the PGF process, shown in Fig.~\ref{fig:scaling-gluon}, that in the
threshold limit, $s\to 4m^2$ or $\eta\to 0$, the results for $k=P$ (solid squares) approach those 
for $k=T$ (open squares). The higher the $Q^2$ the earlier in $\eta\to 0$ this happens. 
Hence, for $\eta\to 0$ the corresponding total partonic spin asymmetry $\sigma_{P,\Pg}/\sigma_{T,\Pg}$
approaches unity. If one recalls the definition of $\sigma_{T,\Pg}$ and $\sigma_{P,\Pg}$
as the sum and difference of cross sections for the two possible relative helicity alignments 
of the photon and the gluon, respectively, this implies that 
only the contribution for equal helicities can contribute near threshold.

Moreover, in this kinematic regime, the NLO corrections strongly dominate the 
behavior of the PGF scaling function. The LO result vanishes in the limit $\eta\to 0$ due 
to the diminishing phase space available, but the NLO contribution derived from
Fig.~\ref{fig:feyn-virt}~(a) diverges as $1/\beta$, which, along with the suppression from
phase space, leads to a constant. 
In addition, inhibited phase space near threshold only allows for
soft gluon radiation originating from diagram \ref{fig:feyn-gluon}~(d).
As is well known, this leads to a large logarithmic enhancements, i.e., powers of $\log \beta$, 
in each order of perturbation theory that are amenable to all-order resummation techniques if necessary,
see, e.g., Refs.~\cite{Laenen:1998kp,Eynck:2000gz,Kawamura:2012cr}.  
We will elaborate on the threshold behavior of the scaling functions 
a bit further at the end of this section and in App.~\ref{app:threshold}, 
where also compact analytical expressions can be found.

Far above threshold, at large $\eta$, the partonic cross sections $\sigma_{P,\Pg}$ and $\sigma_{T,\Pg}$ 
exhibit a very different behavior as can be also inferred from the PGF scaling functions shown in
Fig.~\ref{fig:scaling-gluon}. All unpolarized projections receive large perturbative
corrections, as the NLO contributions $c_{k,\Pg}^{(1)}$ and $\bar{c}_{k,\Pg}^{(1)}$
both approach a constant value, depending on $\xi$, for $\eta\to\infty$ \cite{Laenen:1992zk}.
This behavior can be traced back to real gluon emission with a gluon exchange
in the $t$-channel, i.e., Fig.~\ref{fig:feyn-virt}~(d), which is absent, of course, at the Born
approximation. As in the case of HQ photoproduction \cite{Bojak:1998bd}, such large corrections
are not found for the spin-dependent total partonic cross section $\sigma_{P,\Pg}$. Apparently,
they cancel in the difference of the two possible relative helicity alignments of the
photon and gluon, and both $c_{P,\Pg}^{(1)}$ and  $c_{P,\Pg}^{(0)}$ approach zero as $\eta\to\infty$.
As a consequence, if the hadronic DIS structure functions, obtained as the convolution 
of the scaling functions and the PDFs, see Sec.~\ref{sec:pheno} below, 
predominantly sample the kinematic regime far above threshold,
one has to anticipate very large NLO corrections for the double-spin asymmetry $A_1^{\PQ}$.

\begin{figure}[th!]
\begin{center}
\includegraphics[width=0.48\textwidth]{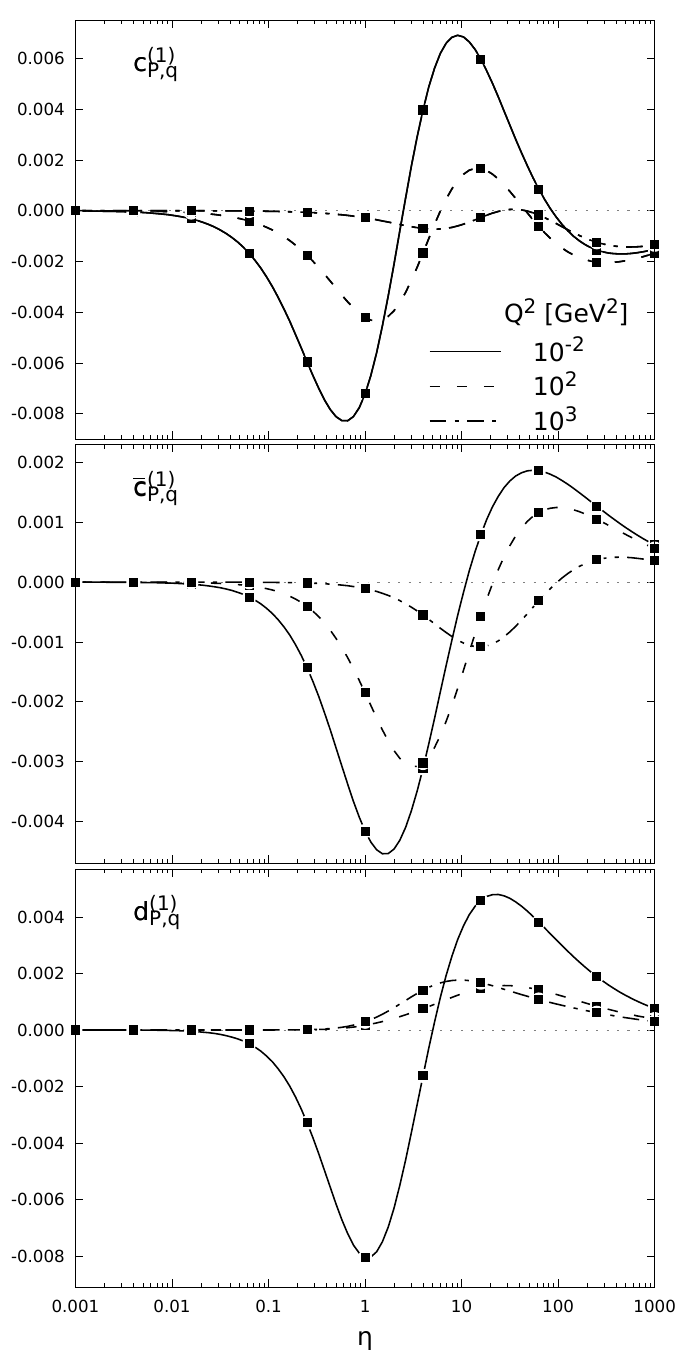}
\end{center}
\vspace*{-0.5cm}
\caption{\label{fig:scaling-quark} 
The top, middle, and lower panel show the light-quark scaling functions 
$c_{P,\Pq}^{(1)}$, $\bar{c}_{P,\Pq}^{(1)}$, and $d_{P,\Pq}^{(1)}$,
respectively, at NLO accuracy for three different values of $Q^2$ 
and fixed $m=\SI{4.75}{\GeV}$.}
\end{figure}
The genuine NLO light-quark scaling functions $c_{k,\Pq}^{(1)}$,
$\bar{c}_{k,\Pq}^{(1)}$, and $d_{k,\Pq}^{(1)}$ for the Bethe-Heitler and Compton processes, 
shown for $k=P$ in the upper, middle, and lower panels of
Fig.~\ref{fig:scaling-quark}, respectively, are numerically much smaller than 
the scaling functions for the PGF mechanism. 
They only exhibit a non-trivial, oscillatory behavior, rapidly decreasing with increasing $\xi$,
in the range $0.1\lesssim \eta\lesssim 100$ and tend to zero both at threshold and 
for asymptotically large values of $\eta$. Nevertheless, depending on the size of
the still only purely constrained gluon helicity distribution $\Delta \Pg$, the contribution
of the light-quark initiated processes to $g_1^{\PQ}$ can be much more significant 
than in the unpolarized case, which is known to be strongly gluon dominated.

Next, we return to the behavior of the gluonic scaling functions $c_{k,\Pg}^{(n)}(\eta,\xi)$ and
$\bar c_{k,\Pg}^{(n)}(\eta,\xi)$ near threshold, i.e., for $\eta\to 0$, or, equivalently, $s\to 4m^2$,
in a bit more detail.
At LO we reproduce the well-known result \cite{Laenen:1992zk}
\begin{eqnarray}
c_{T,\Pg}^{(0),\text{thr}} &=& c_{P,\Pg}^{(0),\text{thr}} =\frac{\pi}{2}\frac{\rho_q}{\rho_q-1} \beta + O(\beta^3)\\
c_{L,\Pg}^{(0),\text{thr}} &=& -\frac{4\pi}{3}\frac{\rho_q^2}{(\rho_q-1)^3}\beta^3 + O(\beta^5)
\end{eqnarray}
that the longitudinal partonic cross section vanishes by two powers of $\beta$ faster than the transverse one, with 
the latter being identical to the polarized $c_{P,\Pg}^{(0),\text{thr}}$.
At NLO accuracy we find
\begin{eqnarray}
\label{eq:thr-approx}
c_{k,\Pg}^{(1),\text{thr}} &=& c_{k,\Pg}^{(0),\text{thr}}
\frac {1} {\pi^2} \Bigg[ C_A \Big( a_k^{(1,2)}\ln^2(\beta) 
                                   + a_k^{(1,1)}\ln(\beta) \nonumber \\
&-& \frac{\pi^2}{16\beta} + a_{k,\tOK}^{(1,0)}\Big) 
+ 2C_F\left(\frac{\pi^2}{16\beta} + a_{k,\tQED}^{(1,0)}\right) \Bigg]\quad \label{eq:cg1threshold}
\end{eqnarray}
with
\begin{eqnarray}
a_T^{(1,2)} &=& a_P^{(1,2)} = a_L^{(1,2)} = 1 \;,\nonumber \\
a_T^{(1,1)} &=& a_P^{(1,1)} = 3\ln(2) - \frac 5 2 \;,\nonumber \\
a_L^{(1,1)} &=& a_T^{(1,1)} - \frac{2}{3}\;.
\end{eqnarray}
The somewhat lengthy expressions for the subleading 
$a_{k,\tOK}^{(1,0)}$ and $a_{k,\tQED}^{(1,0)}$
are given in App.~\ref{app:threshold} along with some brief remarks on how to derive them.
We stress that these subleading coefficients are crucial in order to obtain 
a smooth matching between the threshold approximation in Eq.~(\ref{eq:thr-approx})
and the exact scaling functions $c_{k,\Pg}^{(1)}$ in the threshold region below $\eta\approx 10^{-3}$.
Keeping just the leading-logarithmic (LL), $a_k^{(1,2)}$, and next-to-LL (NLL), $a_k^{(1,1)}$,
approximations in Eq.~(\ref{eq:thr-approx}) is not sufficient for phenomenological applications.
Our results at LL and NLL accuracy for $k=\{T,L\}$ fully agree with the corresponding expressions given in
Ref.~\cite{Laenen:1992zk} except not for $a_L^{(1,1)}$, 
where some piece is missing in Eq.~(5.7) of Ref.~\cite{Laenen:1992zk}. 
We are not aware of a correction of this typo in the literature so far. 
Moreover, the derivation of the results in the threshold limit in Ref.~\cite{Laenen:1992zk} 
contains some but not all of the subleading contributions to 
$a_{k,\tOK}^{(1,0)}$ and $a_{k,\tQED}^{(1,0)}$.
We note that our results for $a_{T,OK/QED}^{(1,0)}$ agree to expressions given in Ref.~\cite{Kawamura:2012cr}, and for $a_{P,OK/QED}^{(1,0)}$ they agree in the photoproduction limit, $Q^2\to 0$, also given for completeness in App.~\ref{app:threshold},
with corresponding expressions given in Ref.~\cite{ref:wilco}.

The threshold limit of the scaling functions $\bar c_{k,\Pg}^{(1)}$ is given by
\begin{equation}
\bar c_{k,\Pg}^{(1),\text{thr}} = c_{k,\Pg}^{(0),\text{thr}} \frac{1}{\pi^2}
     C_A\left(\bar a_k^{(1,1)}\ln(\beta) + \bar a_{k}^{(1,0)}\right)\;,
\end{equation} 
where
\begin{eqnarray}    
\bar a_T^{(1,1)} &=& \bar a_P^{(1,1)} = \bar a_L^{(1,1)} = - \frac 1 2 \;,\nonumber\\
\bar a_T^{(1,0)} &=& \bar a_P^{(1,0)} = -\frac 3 4\ln(2) + \frac 1 2 + 
\frac 1 4 \ln\left(\frac{(1+\chi_q)^2}{2\chi_q}\right) \;,\nonumber \\
\bar a_L^{(1,0)} &=& \bar a_T^{(1,0)} + \frac 1 6\;,
\end{eqnarray}
in agreement with the results at LL given in Ref.~\cite{Laenen:1992zk}.

Finally, the high-energy limit $s\rightarrow \infty$, i.e.\ $\eta\to \infty$, 
our unpolarized results for $c_{T,\Pg}^{(1)}$ and $c_{L,\Pg}^{(1)}$ agree numerically
with the formulae given in Ref.~\cite{Catani:1990eg}. 
Analytic results have been obtained in Refs.~\cite{Buza:1995ie,Buza:1996xr} for all three projections
$k=\{G,\,L,\,P\}$ in the asymptotic limit $Q^2 \gg m^2$.
Our analytic expressions for $\bar c_{k,\Pg}^{(1)}$, $\bar c_{k,\Pq}^{F,(1)}$, and $d_{k,\Pq}^{(1)}$ listed
in App.~\ref{app:scaling-fcts} match with these results. For the other scaling functions
we find perfect numerically agreement for all projections $k$.\\
\section{Phenomenological Applications: HQ DIS Structure Functions and the 
Double-Spin Asymmetry \label{sec:pheno}}
Finally, we turn to some first phenomenological applications of our higher-order corrections.
Since HQ production in longitudinally polarized DIS has not been measured yet, we concentrate
in this paper on inclusive structure functions, in particular, on the charm contribution to
$g_1(x,Q^2)$ that will be accessible at a future EIC \cite{Boer:2011fh} and, most likely, of
significant phenomenological relevance in determining $\Delta \Pg$ at small values of momentum fraction $x$. 
In a forthcoming publication \cite{ref:inprep}, we will study more exclusive observables in helicity-dependent DIS
such as differential distributions in the rapidity and transverse momentum of the produced HQ as well as $\PQ\PaQ$ correlations.

The experimentally accessible HQ contributions to hadronic structure functions 
in terms of the standard DIS variables $x$ and $Q^2$
are related to the total partonic HQ production cross sections $\sigma_{k,j}$ computed up to NLO order accuracy
in Eq.~(\ref{eq:partonic-xsec}) through some kinematic prefactors and a convolution in $z=Q^2/s'$
with appropriate combinations of unpolarized ($k=T,L$) or
polarized ($k=P$) gluon, quark, and antiquark PDFs $f_{k,j}$:
\begin{eqnarray}
F^{\PQ}_{k}(x,Q^2,m^2) &=& \frac{Q^2}{4\pi^2\alpha} \sum_{j=\Pg,\Pq,\Paq}
\int\limits_x^{z_{\max}} \frac{dz}{z} \Bigg[ f_{k,j}\left(\frac{x}{z},\mu_F^2\right) \nonumber \\
&\times& \sigma_{k,j}\left(s,q^2,m^2,\frac{\mu_F^2}{m^2},\frac{\mu_R^2}{m^2}\right) \Bigg]\;,
\label{eq:fq}
\end{eqnarray}
where $z_{max} = Q^2/(4m^2+Q^2)$.
For $k=\{T,L\}$ the usual unpolarized, i.e., helicity-averaged, PDFs appear in (\ref{eq:fq}). 
We note that for $k=P$ the relevant helicity-dependent PDFs $f_{P,j}$ are usually denoted as $\Delta f_j$
in the literature \cite{deFlorian:2008mr,Blumlein:2010rn}, as they refer to 
the difference of densities for the two helicity alignments of the parton spins with
respect to the direction of the nucleon spin.

In terms of the gluon and quark scaling functions introduced in Sec.~\ref{sec:partonic}, 
Eq.~(\ref{eq:fq}) can be written more explicitly as
\begin{widetext}
\begin{eqnarray}
\nonumber
F^{\PQ}_{k}(x,Q^2,m^2) &=& \frac{\alpha_s(\mu_R^2)}{4\,\pi^2} \frac{Q^2}{m^2} 
\int\limits_x^{z_{max}}\frac{dz}{z} \Bigg\{ f_{k,\Pg}\left(\frac{x}{z},\mu_F^2\right) e_H^2 
c^{(0)}_{k,\Pg}(\eta,\xi) \\
\nonumber
&+& 4\pi\alpha_s(\mu_R^2) \Bigg( f_{k,\Pg}\left(\frac{x}{z},\mu_F^2\right) e_H^2 
\left[ c_{k,\Pg}^{(1)}(\eta,\xi) + \bar c_{k,\Pg}^{F,(1)}(\eta,\xi) \ln\left(\frac{\mu_F^2}{m^2}\right)
+ \bar c_{k,\Pg}^{R,(1)}(\eta,\xi) \ln\left(\frac{\mu_R^2}{m^2}\right)\right] \\
&+& \sum_{\Pq}^{n_{lf}} \left[f_{k,\Pq}(x/z,\mu_F^2)+f_{k,\Paq}(x/z,\mu_F^2)\right] \left[ e_H^2
\left( c_{k,\Pq}^{(1)}(\eta,\xi) + \ln\left(\frac{\mu_F^2}{m^2}\right)\bar c_{k,\Pq}^{F,(1)}(\eta,\xi)\right)
+ e_q^2\,  d_{k,\Pq}(\eta,\xi) \right] \Bigg) \Bigg\},\nonumber\\
\label{eq:f2q-hadr}
\end{eqnarray}
\end{widetext}
where the first and second line denotes the LO contribution and the NLO corrections to the PGF process, respectively, and the
third line represents the genuine NLO corrections from the light-(anti)quark-initiated Bethe-Heitler and
Compton processes. In the latter case, the sum includes only the $n_{lf}$ light-quark flavors, e.g.,
$\Pq=\Pqu,\Pqd,\Pqs$ in case of charm production. The partonic variable $\eta=s/(4m^2)-1$ in the scaling functions in
(\ref{eq:f2q-hadr}) is given in terms of $z$ by 
\begin{equation}
\eta= \frac{1-z}{z} \, \frac{Q^2}{4m^2} -1
\end{equation}
for a given HQ mass $m$ and fixed $Q^2$.

The more commonly used DIS HQ structure functions $F_1^{\PQ}$, $F_2^{\PQ}$, $F_L^{\PQ}$, and $g_1^{\PQ}$ are readily expressed in
terms of the three projections $k\in\{T,L,P\}$ in Eq.~(\ref{eq:f2q-hadr}) adopted throughout our calculations:
\begin{eqnarray}
F_{1}^{\PQ}(x,Q^2,m^2) &=& F_{T}^{\PQ}(x,Q^2,m^2)/(2x),\\
F_{2}^{\PQ}(x,Q^2,m^2) &=& F_{T}^{\PQ}(x,Q^2,m^2) + F_{L}^{\PQ}(x,Q^2,m^2),\quad\\
g_1^{\PQ}(x,Q^2,m^2) &=& F_{P}^{\PQ}(x,Q^2,m^2)/(2x)\;,
\end{eqnarray}
i.e., $F_L^{\PQ}=F_2^{\PQ}-2\,x\,F_1^{\PQ}$.
In case of longitudinally polarized lepton and nucleon beams, the experimentally relevant quantity is
the double-spin asymmetry $A_1^{\PQ}$ already introduced in Eq.~(\ref{eq:spin-asym}).

In the remainder of the paper, we perform some first phenomenological studies 
based on our NLO results for charm quark electroproduction.
We will focus on the relevance of the higher order corrections for the inclusive 
structure functions $g_1^{\Pqc}$ and $F_1^{\Pqc}$ and the
corresponding double-spin asymmetry $A_1^{\Pqc}$ in a kinematic range accessible at a future EIC. 
We shall briefly discuss the prospects of further constraining the helicity gluon distribution $f_{P,\Pg}=\Delta \Pg$. 
We leave, however, a detailed impact study of future HQ electroproduction data 
based on a realistic set of pseudo-data for $A_1^{\Pqc}$, that is 
embedded in a global QCD analysis framework for helicity PDFs \cite{Aschenauer:2012ve}, 
to a dedicated future study. 
We shall also estimate the remaining theoretical uncertainties at NLO accuracy due to the choice
of factorization and renormalization scales, as well as the actual value for the charm quark mass
used in the calculations. 

\begin{figure}[bt!]
\begin{center}
\includegraphics[width=0.49\textwidth]{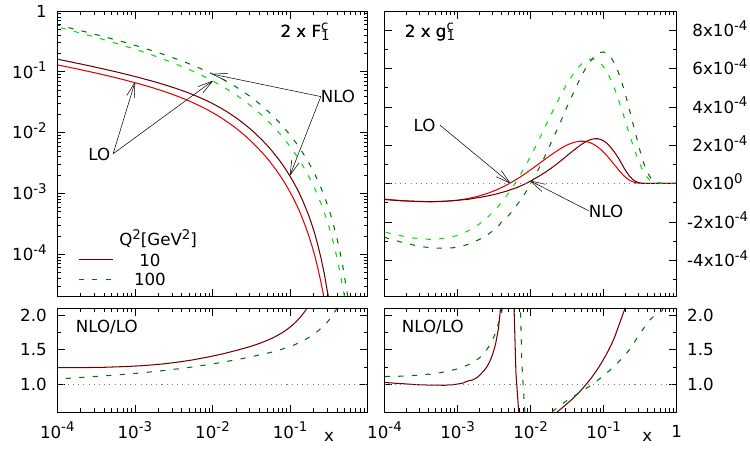}
\end{center}
\vspace*{-0.5cm}
\caption{\label{fig:f1g1} 
The DIS charm structure functions $2x\,F_1^{\Pqc}$ (left-hand-side) 
and $2x\,g_1^{\Pqc}$ (right-hand-side) at LO and NLO accuracy as a function of $x$ 
for two different values of $Q^2$. 
The lower panels show the respective $K$-factors, see text; the result for $g_1^{\Pqc}$ is
difficult to display because of the zero near $x=7\times 10^{-3}$. 
All results were obtained for $m=m_{\Pqc}=\SI{1.5}{\GeV}$ and $\mu_F^2=\mu_R^2=4m^2+Q^2$.}
\end{figure}
\begin{figure*}[th!]
\begin{center}
\includegraphics[width=0.65\textwidth]{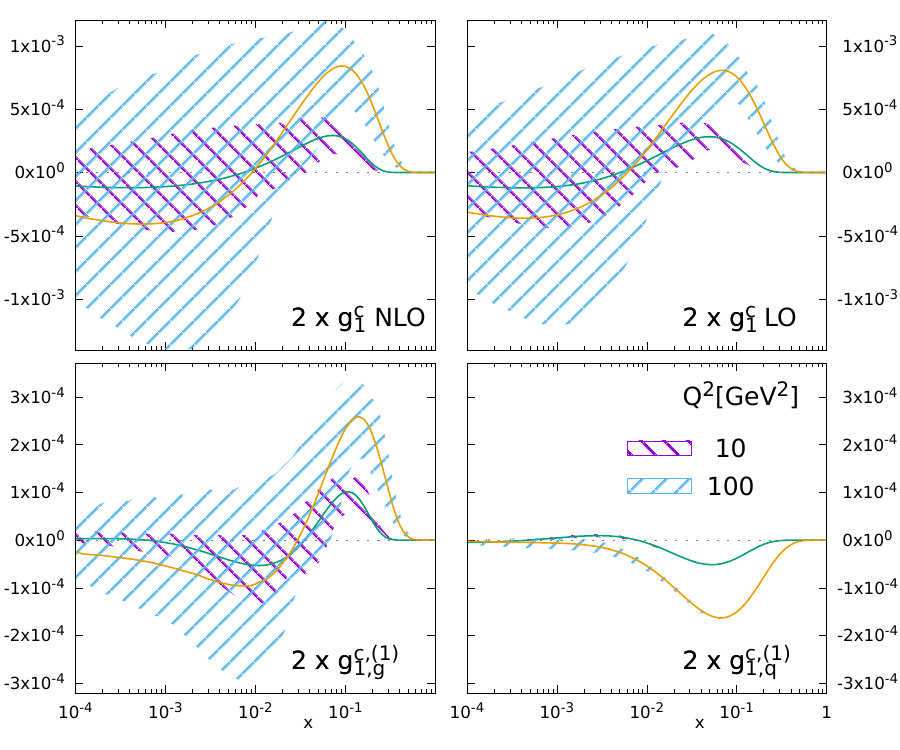}
\end{center}
\vspace*{-0.5cm}
\caption{\label{fig:g1-contr} The shaded bands illustrate the spread in the predictions
for $2x\,g_1^{\Pqc}$ at NLO and LO accuracy (upper row) due to the 
uncertainties of the DSSV helicity PDFs
as a function of $x$ for two different values of $Q^2$. The solid lines refer to
the best fit of DSSV. The lower panels show the gluon and light-(anti)quark
initiated NLO contributions to $g_1^{\Pqc}$, evaluated from the second and third row of Eq.~(\ref{eq:f2q-hadr}),
respectively, and their corresponding spread due to PDF uncertainties.}
\end{figure*}
Unless stated otherwise, we use the DSSV set of helicity PDF \cite{deFlorian:2008mr,deFlorian:2014yva} 
for all our studies of $g_1^{\Pqc}$. Since this set is only available at NLO accuracy, 
we have to evaluate all LO results also with NLO sets of PDFs. Likewise, for all calculations of 
unpolarized quantities, we use the set of NLO PDFs by the MSTW group \cite{Martin:2009iq}, 
which was also adopted in the DSSV global analysis as the
unpolarized reference set in ensuring the positivity limit for helicity PDFs.
To explore the range of expectations for future measurements of $g_1^{\Pqc}$ or $A_1^{\Pqc}$ at an EIC, we also
make use of the uncertainty sets for helicity PDFs provided by DSSV \cite{deFlorian:2014yva}. 
The resulting bands will give a rough estimate of how well
such future experiments have to be performed in order to make an impact on constraining $\Delta \Pg$ further
with inclusive, deep-inelastic charm quark production. For our studies, we assume, that in the kinematic range
covered by an EIC, uncertainties in the unpolarized PDFs are negligible, i.e., the obtained bands 
only reflect our current ignorance of helicity PDFs, in particular, of the gluon density $\Delta \Pg$. 
For the factorization scale and the pole mass of the charm quark
our default choice is $\mu_F^2=\mu_R^2=4m^2+Q^2$ and $m=m_{\Pqc}=\SI{1.5}{\GeV}$, respectively.

Figure~\ref{fig:f1g1} shows the DIS charm structure functions $2x\,F_1^{\Pqc}$ 
and $2x\,g_1^{\Pqc}$ at LO and NLO accuracy as a function of $x$ for two different values of $Q^2$.
The lower panels display the respective $K$-factors, defined, as usual, as the ratio of the NLO and the LO
approximations to Eq.~(\ref{eq:f2q-hadr}). First of all, one notices that $g_1^{\Pqc}$ is significantly smaller in magnitude than
$F_1^{\Pqc}$, which will be even more apparent in the corresponding double-spin asymmetry to be discussed below.
Also, due to its oscillatory behavior, $g_1^{\Pqc}$ has to be displayed on a linear scale, 
and the zeros near $x\simeq 7\times 10^{-3}$ explain the 
kink in the corresponding $K$-factor. 
In the small-$x$ region, the NLO corrections to both $F_1^{\Pqc}$ and $g_1^{\Pqc}$
turn out to be moderate.
In the unpolarized case, the $K$-factor is always larger than unity and decreases for all
values of $x$ with increasing virtuality $Q^2$. There is no such simple systematics for $g_1^{\Pqc}$.
For $x\gtrsim 0.1$, the NLO corrections for $F_1^{\Pqc}$ and $g_1^{\Pqc}$ are very similar and 
both grow rapidly with increasing $x$, i.e., when getting closer to threshold. 
This is readily understood from the behavior of the partonic scaling functions for $\eta\to 0$, 
discussed in Sec.~\ref{sec:partonic}, that becomes more and more relevant at large $x$.

\begin{figure}[thb!]
\begin{center}
\includegraphics[width=0.42\textwidth]{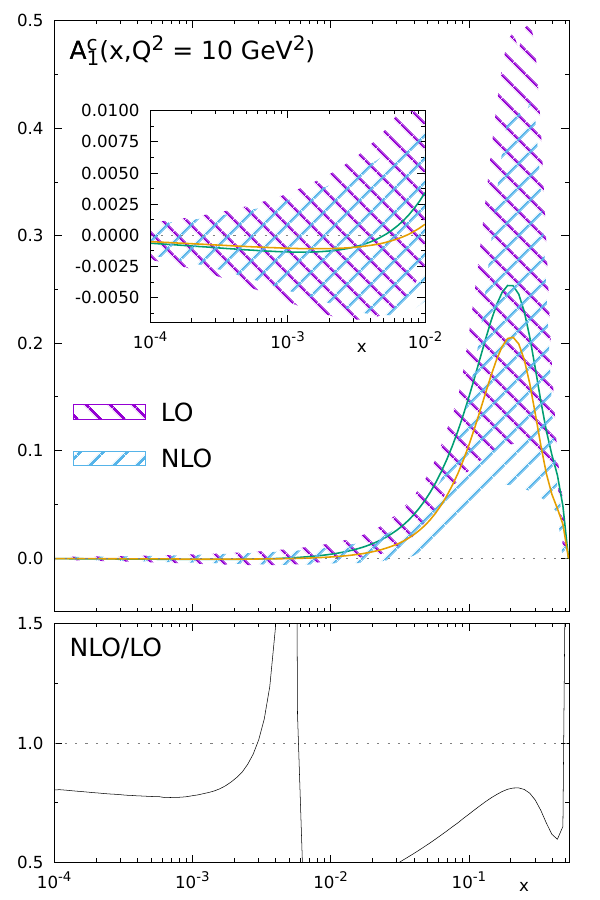}
\end{center}
\vspace*{-0.5cm}
\caption{The spin asymmetry $A_1^{\Pqc}$ for charm quark electroproduction (upper panel) as defined in
Eq.~(\ref{eq:spin-asym}) at LO and NLO accuracy for $Q^2=\SI{10}{\GeV^2}$.
The shaded bands represent the spread in predictions estimated from the uncertainties
of the DSSV set of helicity PDFs. The small inset zooms into the phenomenologically
interesting small-$x$ region.
The lower panel gives the corresponding $K$-factor for the optimum set helicity PDFs from
the DSSV group.
\label{fig:a1c}}
\end{figure}
In Fig.~\ref{fig:g1-contr} we allow for variations of the helicity PDFs within the uncertainty
bands estimated by the DSSV group \cite{deFlorian:2014yva}. As can be seen, the resulting spread in
$g_1^{\Pqc}$ (shaded bands) is very large, in particular, for small values of momentum fraction $x$.
In the lower panels, we show separately the gluon and light-quark induced NLO contributions,
$g_{1,\Pg}^{\Pqc,(1)}$ and $g_{1,\Pq}^{\Pqc,(1)}$, evaluated from the second and third row of Eq.~(\ref{eq:f2q-hadr})
respectively. As expected, the uncertainties in
the poorly constrained gluon helicity PDF cause a much bigger variation in $g_1^{\Pqc}$ than those
stemming from all the light quark PDFs together.
It is also worth noticing, that for the optimum set of DSSV (solid lines) the light-quark
induced processes are roughly of the same size as the NLO contribution from PGF.
This is very much at variance of what is known in the unpolarized case, which is strongly gluon
dominated \cite{Laenen:1992zk}. Present uncertainties in $\Delta \Pg$ still allow, however,
for a PGF dominance also for $g_1^{\Pqc}$.
In the region around $x\simeq 0.1$, where the PGF process is positive, light quarks contribute
to $g_1^{\Pqc}$ with the opposite sign, which will diminish the experimentally  relevant
double-spin asymmetry $A_1^{\Pqc}$ in this kinematic regime.

\begin{figure}[th!]
\begin{center}
\includegraphics[width=0.41\textwidth]{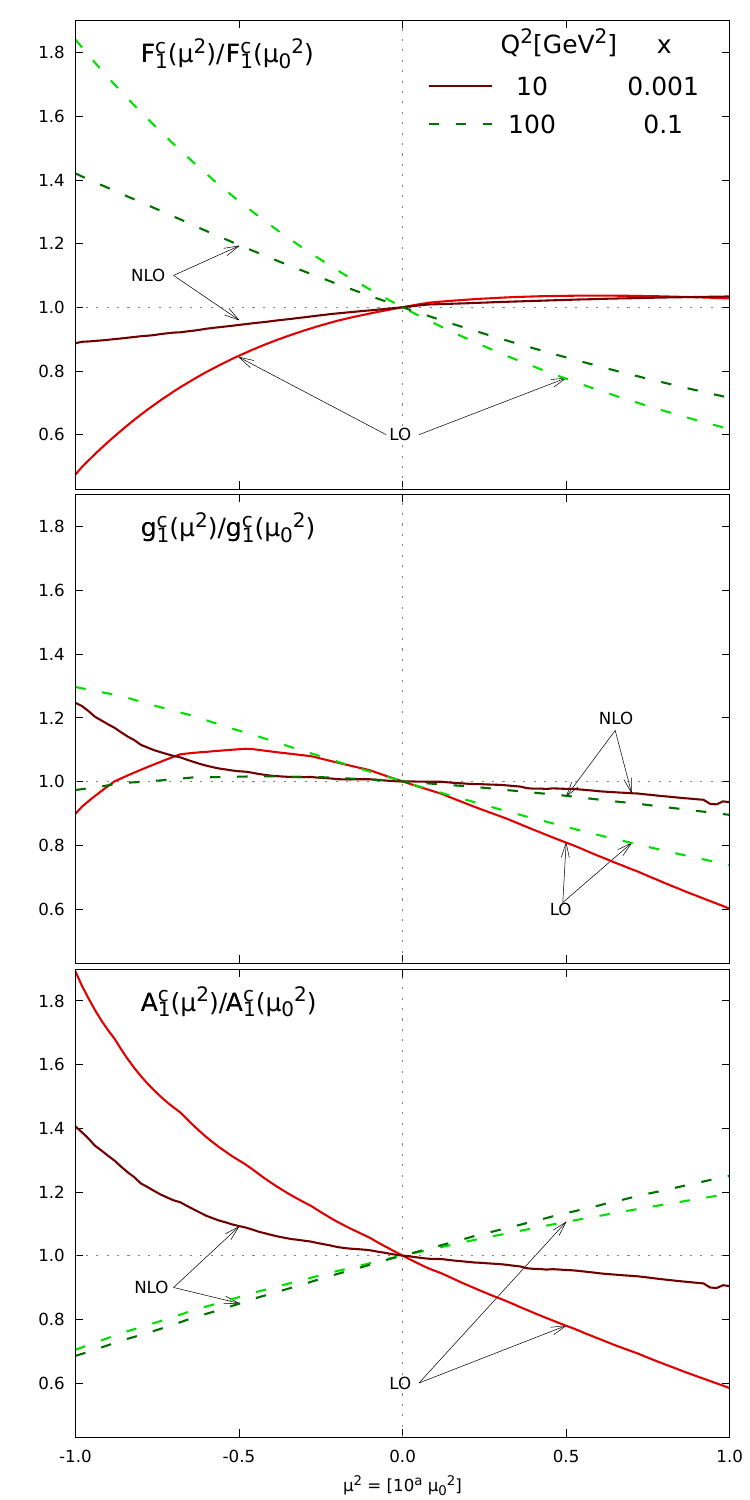}
\end{center}
\vspace*{-0.5cm}
\caption{The scale dependence of $F_1^{\Pqc}$, $g_1^{\Pqc}$, and $A_1^{\Pqc}$ at LO and NLO accuracy
for two different pairs of $x$ and $Q^2$ values
for $\mu^2=\mu_F^2=\mu_R^2$ in the range $\mu_0^2/10\le\mu^2\le 10\mu_0^2$,
where $\mu_0^2=4m^2+Q^2$ is our default choice of scale.
In each panel, the results are normalized to the ones obtained for $\mu^2=\mu_0^2$.
\label{fig:a1c-mu2}}
\end{figure}
$A_1^{\Pqc}$, defined in Eq.~(\ref{eq:spin-asym}), is shown in Fig.~\ref{fig:a1c} 
as a function of $x$ at LO and NLO accuracy for $Q^2=\SI{10}{\GeV^2}$. 
The range around $x\simeq 10^{-3}$ is expected to be accessible at a
future EIC at this particular value of $Q^2$ \cite{Boer:2011fh}.
Again, the shaded bands reflect the estimates of uncertainties for the DSSV set of helicity PDFs. Since
the spin asymmetry, like $g_1^{\Pqc}$, changes sign, it has to be displayed on a linear scale. 
To better visualize the behavior of $A_1^{\Pqc}$ in the regime of phenomenological interest to an EIC, 
the inset in the upper panel of Fig.~\ref{fig:a1c} zooms into the small $x$-region.
At $x\simeq 10^{-3}$ the spread in $A_1^{\Pqc}$ due to current uncertainties in $\Delta \Pg$ 
ranges from about $-0.005$ to $+0.004$, which implies that a future measurement of
$A_1^{\Pqc}$ at an EIC should aim for an experimental precision at the level of $\mathcal{O}(10^{-3})$ 
or better in order to make an impact in further constraining helicity PDFs from HQ electroproduction.
The lower panel of Fig.~\ref{fig:a1c} shows the $K$-factor, which, at $Q^2=\SI{10}{\GeV^2}$,
is always smaller than unity but roughly constant for $10^{-4}\lesssim x \lesssim 10^{-3}$.
This finding is readily understood from the individual $K$-factors for the DIS structure functions $F_1^{\Pqc}$ and $g_1^{\Pqc}$, 
presented in Fig.~\ref{fig:f1g1}. As for $g_1^{\Pqc}$, the node at $x\sim 5\times 10^{-3}$ in $A_1^{\Pqc}$
explains the kink in the $K$-factor.

\begin{figure*}[th!]
\begin{center}
\includegraphics[width=0.42\textwidth]{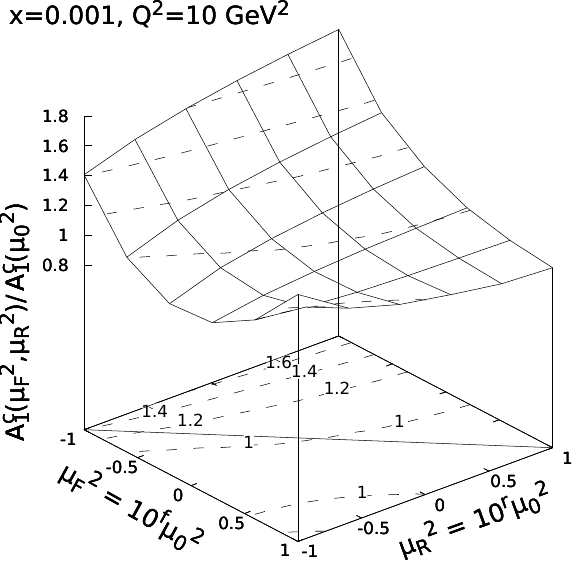}
\includegraphics[width=0.42\textwidth]{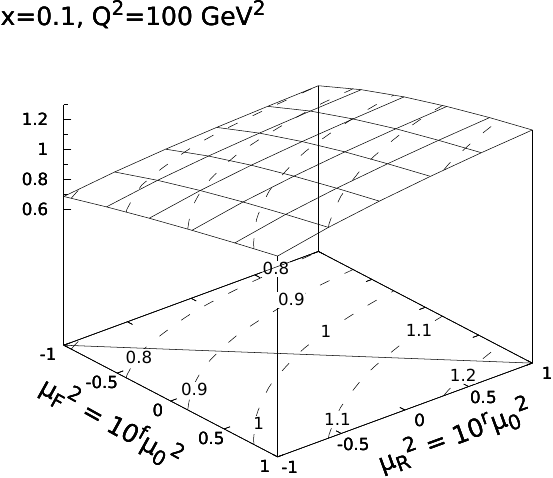}
\end{center}
\vspace*{-0.5cm}
\caption{Similar to Fig.~\ref{fig:a1c-mu2} but now for independent variations of
$\mu_F^2$ (on the ''f-axis'') and $\mu_R^2$ (''r-axis'') in the 
computation of the double-spin asymmetry $A_1^{\Pqc}$ at NLO accuracy for 
$x=10^{-3}$ and $Q^2=\SI{10}{\GeV^2}$ (l.h.s.) and
$x=0.1$ and $Q^2=\SI{100}{\GeV^2}$ (r.h.s.). Again,
all results are normalized to the one obtained for
$\mu_F^2=\mu_R^2=\mu_0^2 = 4m^2+Q^2$. The base of each plot shows lines of constant ratio (dashed lines), 
and the solid line indicates the choice $\mu_F^2=\mu_R^2$ adopted in Fig.~\ref{fig:a1c-mu2}.
\label{fig:a1c-mufmur}}
\end{figure*}
Next, we turn to some estimates of residual theoretical uncertainties in HQ quark electroproduction
from variations of the factorization and renormalization scales and the charm quark mass.
Figure~\ref{fig:a1c-mu2} shows the dependence of $F_1^{\Pqc}$, $g_1^{\Pqc}$, and $A_1^{\Pqc}$ 
on simultaneous variations of scales, $\mu^2=\mu_F^2=\mu_R^2$, 
in the broad range $\mu_0^2/10\le\mu^2\le 10\mu_0^2$ for two
pairs of $x$ and $Q^2$ values accessible at an EIC. All results are normalized to those obtained
with our default choice of scale, $\mu_0^2=4m^2+Q^2$, used in Figs.~\ref{fig:f1g1}-\ref{fig:a1c}.
As can be seen, the DIS structure functions $F_1^{\Pqc}$ and $g_1^{\Pqc}$ exhibit significantly
smaller variations with scale at NLO accuracy than the corresponding LO results. In general,
variations in $g_1^{\Pqc}$ for $\mu^2\ll \mu_0^2$ turn out to be smaller than those for $F_1^{\Pqc}$, but
differences between LO and NLO results are more pronounced for $\mu^2\gg \mu_0^2$.
Due to the different behavior of $F_1^{\Pqc}$ and $g_1^{\Pqc}$ both with variations of scales and concerning
higher order corrections, the scale dependence of the experimentally relevant
double-spin asymmetry $A_1^{\Pqc}$ is non-trivial. Again, we find considerably more stable results
at NLO accuracy for small values of $x$, but little or no improvement for $x\simeq0.1$.  

In Fig.~\ref{fig:a1c-mufmur} we perform similar changes of scale for
$A_1^{\Pqc}$ at NLO accuracy for the same two pairs of $x$ and $Q^2$ values 
used in Fig.~\ref{fig:a1c-mu2} but now allowing for independent variations of $\mu_F$ and $\mu_R$.
To guide the eye, we show contour lines of constant ratio (dashed lines),
evaluated as before with respect to our default choice of scale, also at the
base of the plot. The solid line indicates the choice $\mu_F^2=\mu_R^2$ used
in Fig.~\ref{fig:a1c-mu2}. As one can anticipate, choosing $\mu_F$ and $\mu_R$ very differently
introduces large logarithms $\propto \ln(\mu_F^2/\mu_R^2)$ in the
partonic cross sections that can lead to large variations of $A_1^{\Pqc}$ as
compared to the default choice $\mu_F=\mu_R=\mu_0$.
This can be observed, in particular, for the smaller value of $x=10^{-3}$, shown on the
l.h.s.\ of Fig.~\ref{fig:a1c-mufmur}, whereas the results for $x=0.1$ and
$Q^2=\SI{100}{\GeV^2}$ (r.h.s.) turn out to be considerably more stable.
It is interesting to notice, however, that for $x=10^{-3}$ the line at the base of the plot
indicating the choice $\mu_F=\mu_R$ is fairly close to the contour line for 1, whereas the
corresponding line for $x=0.1$ crosses the contour lines almost perpendicular. 
The lesson is, that the theoretical uncertainties for $A_1^{\Pqc}$
from scale variations have a strong dependence on the kinematics probed, i.e., 
the actual values of $x$ and $Q^2$, and cannot be simply approximated or estimated.
Most importantly, scale variations do not cancel in the double-spin asymmetry as
one may naively expect for a ratio of two cross sections.
The main reason is the different behavior of the numerator and denominator of
$A_1^{\Pqc}$ on variations of the scale and, in addition, also on the NLO corrections.

\begin{figure}[t!]
\begin{center}
\includegraphics[width=0.48\textwidth]{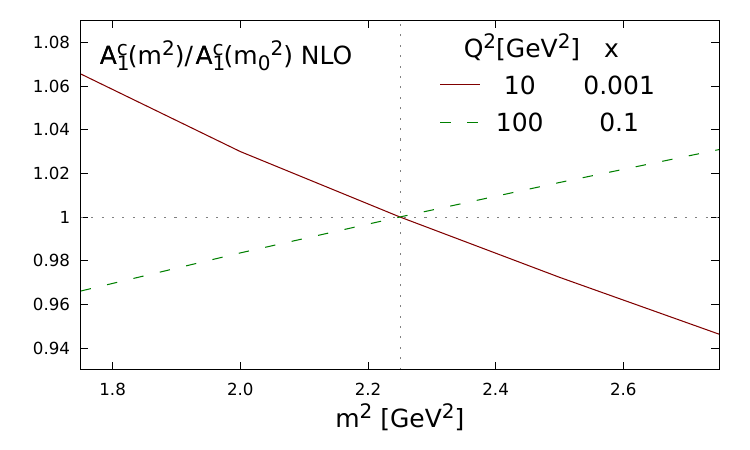}
\end{center}
\vspace*{-0.5cm}
\caption{Dependence of the double-spin asymmetry $A_1^{\Pqc}$ for charm electroproduction 
on the choice of $m^2$ for two pairs of $x$ and $Q^2$ values. All results
are normalized to the one obtained for our default choice of the charm quark pole mass 
$m=m_0=\SI{1.5}{\GeV}$ (vertical dotted line).
\label{fig:a1c-m2}}
\end{figure}
Finally, the actual value for HQ pole mass $m$ used in the calculation of the DIS structure functions
can also have an impact on the experimentally relevant double-spin asymmetry and, as for the variations of
$\mu_F$ and $\mu_R$ discussed above, might not cancel in the ratio (\ref{eq:spin-asym}).
These two points are addressed in Fig.~\ref{fig:a1c-m2} for the same two pairs of $x$ and $Q^2$ values 
used in Figs.~\ref{fig:a1c-mu2} and \ref{fig:a1c-mufmur} for variations of scales.
Here, we show the ratio of $A_1^{\Pqc}$ computed at NLO accuracy for a range of charm quark masses $m$ 
relative to the results obtained for our default choice $m=m_{\Pqc}=\SI{1.5}{\GeV}$.
The theoretical ambiguities introduced by these variations of $m_{\Pqc}$ are significantly smaller
than those found for variations of the factorization and renormalization scales
shown in Figs.~\ref{fig:a1c-mu2} and \ref{fig:a1c-mufmur}. 
For the two selected kinematic points the variations due to $m$ rarely exceed $\pm 5\%$.
Nevertheless, the choice of different masses $m$ in the calculation of DIS structure functions
does not cancel in $A_1^{\Pqc}$  and, as for the variations of scales above, 
has a nontrivial dependence on the selected values for $x$ and $Q^2$. 
For our selection of kinematic points it even leads to different slopes with respect to 
variations of $m$. We note, that in the future, for phenomenological studies of
very precise EIC data it might be advantageous to replace the traditionally used pole mass 
by the running mass definition in the $\overline{\text{MS}}$ scheme in calculations of $g_1^{\PQ}$. 
In case of unpolarized DIS this transformation was first applied in Ref.~\cite{Alekhin:2010sv} and
demonstrated to lead to reduced theoretical uncertainties due to variations of
$\mu_F$ and $\mu_R$.

\section{Summary and Outlook \label{sec:summary}}
In this paper we have completed the suite of NLO QCD calculations of heavy flavor production
with longitudinally polarized beams and targets by performing the first computation 
of the full NLO QCD corrections for single-inclusive heavy flavor
production in helicity-dependent deep-inelastic scattering.
All results were derived with largely analytical methods and retain the full dependence
on the heavy quark's mass. Whenever feasible, compact analytic expressions have been presented
for the total partonic cross sections in terms of heavy quark scaling functions, which are
required to compute the heavy flavor contribution to the helicity-dependent DIS structure function $g_1$.

As a byproduct, and as an important crosscheck, we have re-derived all known results for unpolarized
heavy quark electroproduction. In addition, we have verified our results against the known limit
of photoproduction in both the unpolarized and the polarized case. 
The behavior of the scaling functions was investigated also in various other important limits, 
namely for asymptotically large photon virtualities and close to threshold. In the latter case, 
the leading and next-to-leading logarithmic terms have been verified against known results.
Moreover, analytic expressions for the phenomenological important subleading coefficients were presented
and compared to the literature if available.

We believe that our results are particularly timely and important in view of the ongoing planning 
process towards the realization of a future electron-ion collider, where longitudinally polarized 
DIS will be studied with unprecedented precision even in the so far 
unexplored kinematic regime  of small momentum fractions $x$. Here, for the first time, 
the charm quark contribution to the structure function
$g_1$ will be experimentally accessible and, most likely, of significant phenomenological relevance
when analyzing data in terms of helicity-dependent parton densities, in particular, due to its direct
sensitivity to the gluon distribution already at the Born approximation.
All future global QCD analyses will be aiming mainly at a much improved extraction of the elusive gluon helicity 
density in the small $x$ regime. 

Therefore, and as a first phenomenological application,
we provided some numerical estimates for the charm contribution to $g_1$ and the 
experimentally relevant double-spin asymmetry in the kinematic domain accessible to the planned EIC.
We have demonstrated the sensitivity to the polarized gluon distribution and commented on
the required experimental precision for such measurements at an EIC.
In addition, to get an idea of the residual theoretical uncertainties inherent to the 
polarized electroproduction of heavy quarks at NLO accuracy, 
the dependence of our results on variations of the unphysical factorization and renormalization scales, 
as well as of the choice of charm quark mass, have been studied.
Most importantly, neither these sources of theoretical ambiguities nor the
NLO corrections themselves cancel in the double-spin asymmetry as one may naively expect. 
In general, we find a nontrivial dependence on all these effects on the actual DIS kinematics under consideration.
Adopting the full NLO expressions without approximation in future global QCD analyses of
helicity-dependent parton densities is indispensable.

Further phenomenological applications and extensions of our NLO results are currently under
investigation. In order to allow for flexible experimental cuts it would be advantageous to
combine our NLO matrix elements with a Monte-Carlo sampling of the phase space,
following similar calculations already available in the unpolarized case.
This will enable us to study not only single-inclusive but also exclusive distributions
and correlations of the produced heavy quark pair. It would be interesting to study 
to what extent such observables would help to determine helicity parton densities,
in particular, the gluon density, more precisely at an EIC.

\section*{Acknowledgments}
We are grateful to W.\ Vogelsang for helpful discussions and comments.
This work is supported in part by 
the Bundesministerium f\"{u}r Bildung und Forschung (BMBF) under 
grant no.\ 05P15VTCA1.

\appendix
\section{Analytic Expressions for the Partonic Scaling Functions \label{app:scaling-fcts}}
In this Appendix, we give, whenever possible, compact analytic results for the 
partonic scaling functions presented in Sec.~\ref{sec:partonic}.
Unfortunately, the expressions for $c_{k,j}^{(1)}$, $j=\{\Pg,\,\Pq\}$ are too complex (so far) to be 
presented here but partial results are available upon request from the authors. 

First, we introduce some auxiliary functions $h_{1,2,3}(\chi,\chi_q)$ with
$\chi$ and $\chi_q$ defined in Eq.~(\ref{eq:partonic-var}):
\begin{eqnarray}
\nonumber
h_1(\chi,\chi_q) &=& -\zeta(2)-2\DiLog(-\chi )+
                     \DiLog\left(\frac{1-\chi_q}{1+\chi}\right) \\                     
\nonumber
&+&\DiLog\left(-\frac{1-\chi_q}{(1+\chi )\chi_q}\right)
   -\DiLog\left(\frac{\chi (1-\chi_q)}{1+\chi}\right) \\
&-&\DiLog\left(-\frac{\chi (1-\chi_q)}{(1+\chi)\chi_q}\right)+\frac 1 2 \ln^2 (\chi )\nonumber\\
&+&\ln (\chi )\left[\ln(\chi_q)-\ln (\chi +\chi_q)-\ln (1+\chi  \chi_q) \right],\nonumber\\\\
h_2(\chi,\chi_q) &=& - \zeta(2)+2\DiLog(\chi)+2\DiLog(-\chi)-\frac 1 2 \ln(\chi)\nonumber\\
&-&\ln(\chi)\left[\ln(\chi_q)-\ln(\chi +\chi_q)-\ln(1+\chi\chi_q)\right],\nonumber\\\\
h_3(\chi,\chi_q) &=& \ln (1-\chi )+ \ln (1+\chi ) - \frac{1}{2} \left[\ln (\chi )-\ln (\chi_q) \right. \nonumber \\
&+&\left.\ln (1+\chi  \chi_q)+\ln (\chi +\chi_q)\right]\;.
\end{eqnarray}
Instead of $\eta=1/\rho-1$ and $\xi=-q^2/m^2$, the gluonic scaling functions, $\bar c_{k,\Pg}^{(1)}$, 
and the factorization scale dependence of the quark scaling functions, $\bar c_{k,\Pq}^{F,(1)}$, given in
App.~\ref{app:gluonic} and \ref{app:quarkfact}, respectively, below are most conveniently expressed in terms
of the variables $\beta$, $\beta_q$, $\rho$, and $\rho_q$ defined in Eq.~(\ref{eq:partonic-var}).
Of course, all these variables can be readily re-expressed in terms of $\eta$ and $\xi$ if needed.

Compact analytical expressions can be also found for the light-quark scaling functions $d_{k,\Pq}^{(1)}$,
i.e., the Compton process, for all projections $k$. They are listed in App.~\ref{app:quark-d}. 
Our results, when re-expressed in terms of $\eta$ and $\xi$, 
match exactly with those given in Ref.~\cite{Blumlein:2016xcy} for all $k$.
We note again that $d_{P,\Pq}^{(1)}$ is the only coefficient function 
for longitudinal polarization at NLO accuracy
that was known prior to the calculations presented in this paper.

Again, we find that it is more convenient to express $d_{k,\Pq}^{(1)}(\eta,\xi)$
in terms of auxiliary variables to arrive at very compact expressions.
To this end, we introduce an additional set of variables
$\rho'$, $\beta'$, and $\chi'$, 
\begin{eqnarray}
\label{eq:quark-variables}
0 \leq \rho' &=& \frac{4m^2}{s'} \leq \frac{\rho_q}{\rho_q-1} \leq 1\;, \nonumber \\
0 \leq \beta' &=& \sqrt{1-\rho'} \leq \frac 1 {\beta_q}\leq 1\;, \nonumber \\
0 \leq \chi' &=& \frac{1-\beta'}{1+\beta'} \leq \chi_q \leq 1\;,
\end{eqnarray}
which obey the additional inequalities
\begin{equation}
0\leq \rho' \leq \rho \leq 1\;,\;\; 
0\leq \beta \leq \beta' \leq 1\;,\;\;
0\leq \chi'\leq \chi\leq 1
\end{equation}
with respect to $\rho$, $\beta$, and $\chi$
given in Eq.~(\ref{eq:partonic-var}). 
Again, the new variables are readily re-expressed in terms of $\eta$ and $\xi$.
We also define another auxiliary function which reads
\begin{eqnarray}
h_4(\chi,\chi') &=& \DiLog\left(\frac {1+\chi'}{1+\chi}\right)- \DiLog\left(\chi \frac{1+\chi'}{1+\chi}\right) \nonumber\\
 &-& \DiLog\left(\chi'\frac{1+\chi}{1+\chi'}\right)+ \DiLog\left(\frac{\chi'(1+\chi)}{\chi(1+\chi')} \right)\nonumber\\
 &+& \frac1 2 \ln^2 (\chi ) +  \ln (\chi ) \Big[ \ln (1+\chi )+\ln (1+\chi')  \nonumber \\
 &-& \ln (\chi -\chi')-\ln (1-\chi  \chi')\Big]\;.
\end{eqnarray}

We note that the three sets of $\rho$-, $\beta$- and $\chi$-type variables 
defined in Eqs.~(\ref{eq:partonic-var}) and (\ref{eq:quark-variables}),
which we use to present our results for the coefficient functions, 
are linked by the simple, reciprocal relation 
\begin{equation}
\label{eq:rho-relation}
\frac {1}{\rho'} = \frac 1 \rho - \frac 1 {\rho_q}
\end{equation}
that follows from $s' = s-q^2$ and leads to rather
nontrivial relations for the corresponding $\beta$- and $\chi$-type variables
due to radicals like $\beta=\sqrt{1-\rho}$.

It might be illuminating to elaborate a bit more on usefulness of the
$\chi$-type variables. As was noted in Ref.~\cite{Czakon:2008ii} in case of
unpolarized hadroproduction of HQs, the introduction of
$\chi$ maps the three types of high-energy, threshold, unphysical singularities
of the relevant partonic variables, in their case $s$ and $m$, 
onto the points $\chi=(0,1,-1)$, which allows one to express the partonic results
almost entirely in terms of harmonic polylogarithms \cite{Remiddi:1999ew}.
In our calculation, the analytic structure of the results 
is further complicated by the appearance of the additional scale $q^2$,
which explains the need for more than one set of variables.
We find that the most compact analytic expressions for the partonic
coefficient functions $\bar c_{k,\Pg}^{F/R,(1)}$ and $\bar c_{k,\Pq}^{F,(1)}$
given in App.~\ref{app:gluonic} and \ref{app:quarkfact}, respectively,
are obtained in terms of the variables $\chi$ and $\chi_q$, whereas
for $d_{k,\Pq}^{(1)}$, listed in App.~\ref{app:quark-d}, the sets 
containing $\chi$ and $\chi'$ are more appropriate.

As was already mentioned in Sec.~\ref{sec:partonic},  
analytic results have been obtained in the literature for all three projections
in the asymptotic limit $Q^2 \gg m^2$,
$k=\{G,\,L\}$ in Ref.~\cite{Buza:1995ie} and $k=P$ in Ref.~\cite{Buza:1996xr}. 
Our analytic expressions for
$\bar c_{k,\Pg}^{(1)}$, $\bar c_{k,\Pq}^{F,(1)}$, and $d_{k,\Pq}^{(1)}$ given
in App.~\ref{app:gluonic} -- \ref{app:quark-d} below match with the corresponding
results in \cite{Buza:1995ie,Buza:1996xr}.  
In case of $c_{k,\Pg}^{(1)}$ and $c_{k,\Pq}^{(1)}$, we have checked numerically,
that our expressions agree with those given in \cite{Buza:1995ie,Buza:1996xr} for
all $k$.

\subsection{Gluonic Scaling Functions $\bar c_{k,\Pg}^{(1)}$ \label{app:gluonic}}
For the three projections $k=\{G,\,L,\,P\}$ we find
\begin{widetext}
\begin{eqnarray}
\nonumber
\bar c_{G,\Pg}^{(1)}(\eta,\xi) &=& \frac{\rho_q}{8\pi(\rho_q-1)(\rho_q-\rho)^3} \;
    \Bigg\{ 3 \rho  (\rho -\rho_q) (-1+\rho_q) \left[-\rho_q+\rho  (5+2 \rho_q)\right] h_1(\chi,\chi_q)  \nonumber\\
 &+& \frac{3}{2} \rho (-1+\rho_q) \left[ -2 \rho_q^2-2 \rho  \rho_q^2+\rho ^2 (-2+\rho_q (2+\rho_q))\right] h_2(\chi,\chi_q) 
  + 6 \beta  \rho  (-1+\rho_q) \left(\rho ^2+(1+\rho ) \rho_q^2\right) h_3(\chi,\chi_q)\nonumber\\
 &+& \frac{1}{8} \beta  \rho_q \left[ 4 (4-7 \rho_q) \rho_q^2+2 \rho \rho_q (-31+43 \rho_q)
     +\rho ^2 (200-\rho_q (117+95 \rho_q))\right] \nonumber\\
 &+& \frac{1}{16} \rho  (-1+\rho_q) \left[ 48 \rho_q^2-48 \rho  \rho_q (5+2 \rho_q)
      +\rho ^2 (-8+\rho_q (96+59 \rho_q))\right] \ln (\chi )\nonumber\\
 &+& \frac{(\rho -\rho_q)^2}{4 \beta_q} \left[ (-2+\rho_q) \rho_q (-4+7 \rho_q)+\rho  \left(-8+11 \rho_q^2\right)\right] 
     \ln \left(\frac{\chi +\chi_q}{1+\chi  \chi_q}\right) \Bigg\}\;,\\
\bar c_{L,\Pg}^{(1)}(\eta,\xi) &=& \frac{\rho_q}{8\pi(\rho_q-1)(\rho_q-\rho)^3} \;
      \Bigg\{ -6 \rho ^3 (-1+\rho_q) \rho_q \; h_2(\chi,\chi_q) 
               +24  \beta  \rho ^2 (-1+\rho_q) \rho_q \; h_3(\chi,\chi_q) \nonumber\\
 &+& \beta  \rho_q \left(36 \rho ^2-35 \rho ^2 \rho_q-2 (1+\rho ) \rho_q^2+3 \rho_q^3\right) 
     - 4 \rho ^2 (-1+\rho_q) \left[ 3 \rho_q+\rho  (-6+5 \rho_q) \right] \ln (\chi )  \nonumber\\
 &+& \frac{(\rho -\rho_q)}{2 \beta_q} \left[ 2 \rho  \rho_q^2+\rho_q^3 (-4+3 \rho_q)
     +\rho ^2 (-6+5 \rho_q) (-8+9 \rho_q)\right] \ln \left(\frac{\chi +\chi_q}{1+\chi \chi_q}\right)\Bigg\}\;,\\
\bar c_{P,\Pg}^{(1)}(\eta,\xi) &=& \frac{\rho_q}{8\pi(\rho_q-1)(\rho_q-\rho)^3} \; 
       \Bigg\{ 6 \rho  (-1+\rho_q) \left(2 \rho ^2-3 \rho  \rho_q+\rho_q^2\right) h_1(\chi,\chi_q) \nonumber\\
 &+& 3 \rho  (-1+\rho_q) \left(-\rho ^2+\rho_q^2\right)\; h_2(\chi,\chi_q) 
     + 6 \beta  \rho  (\rho -\rho_q) (-1+\rho_q) (\rho +3 \rho_q)\; h_3(\chi,\chi_q) \nonumber\\
 &-& 48 \beta  \rho  (\rho -\rho_q) (-1+\rho_q) \rho_q 
        -\frac{3}{2} \rho  (\rho -\rho_q) (-1+\rho_q) \left[11 \rho_q+\rho  (-13+4 \rho_q)\right] \ln (\chi )  \nonumber\\
 &-& 6 \beta_q \rho  (-1+\rho_q) \left(3 \rho ^2-4 \rho  \rho_q+\rho_q^2\right) \ln \left(\frac{\chi +\chi_q}{1+\chi  \chi_q}\right) 
   \Bigg\}\;.
\end{eqnarray}
\end{widetext}
The decomposition into the factorization part $\bar c_{k,\Pg}^{F,(1)}$ and the renormalization part $\bar c_{k,\Pg}^{R,(1)}$ has been defined in Eqs. (\ref{eq:cgBarR}) and (\ref{eq:bar-fr-sum}).

\subsection{Light-Quark Scaling Functions $\bar c_{k,\Pq}^{F,(1)}$ \label{app:quarkfact}}
The light-quark scaling functions appear for the first time at NLO accuracy and, hence, 
only carry a dependence on the factorization scale, which reads for the three
projections $k=\{G,\,L,\,P\}$:
\begin{widetext}
\begin{eqnarray}
\bar c_{G,\Pq}^{F,(1)}(\eta,\xi) &=& \frac{\rho_q}{36\pi(\rho_q-1)(\rho_q-\rho)^3} \Bigg\{ 
       3 \rho  (\rho -\rho_q) (-1+\rho_q) \left[ -2 \rho_q+\rho  (4+\rho_q)\right]\; h_1(\chi,\chi_q) \nonumber\\
 &+& \beta  \rho_q \left[ (4-7 \rho_q) \rho_q^2+2 \rho  \rho_q (-7+10 \rho_q)+\rho ^2 (14-\rho_q (15+2 \rho_q))\right] \nonumber\\
 &+& \rho  (-1+\rho_q) \left[ -3 \rho  \rho_q+3 \rho_q^2+\rho ^2 \left(-4+\rho_q^2\right)\right] \ln(\chi)\nonumber\\
 &+& \frac{(\rho -\rho_q)^2}{2 \beta_q}\left[ (-2+\rho_q) \rho_q (-4+7 \rho_q)+\rho  \left(-8+11 \rho_q^2\right)\right] 
     \ln \left(\frac{\chi +\chi_q}{1+\chi  \chi_q}\right)\Bigg\}\;,
\end{eqnarray}
\begin{eqnarray}
\bar c_{L,\Pq}^{F,(1)}(\eta,\xi) &=& \frac{\rho_q}{18\pi(\rho_q-1)(\rho-\rho_q)^3} \Bigg\{
 \beta  \rho_q \left(2 \rho  \rho_q^2+(2-3 \rho_q) \rho_q^2+\rho ^2 (-6+5 \rho_q)\right) %
 + 2 \rho ^2 (-1+\rho_q) (\rho  (-3+\rho_q)+3 \rho_q) \ln(\chi)\nonumber\\
 &+& \frac{(-\rho +\rho_q) }{2 \beta_q}\left(2 \rho  \rho_q^2+\rho_q^3 
     (-4+3 \rho_q)+\rho ^2 (12+\rho_q (-22+9 \rho_q))\right) \ln \left(\frac{\chi +\chi_q}{1+\chi  \chi_q}\right)\Bigg\} \;,\\
\end{eqnarray}
\begin{eqnarray}
\bar c_{P,\Pq}^{F,(1)}(\eta,\xi) &=& \frac{\rho_q \rho }{6\pi(\rho_q-1)(\rho_q-\rho)^3} \Bigg\{ 
     (-1+\rho_q) \left(2 \rho ^2-3 \rho  \rho_q+\rho_q^2\right) h_1(\chi,\chi_q) 
      - 6 \beta  (\rho -\rho_q) (-1+\rho_q) \rho_q \nonumber\\
 &-& \frac{1}{2} (\rho -\rho_q) (-1+\rho_q) (\rho  (-6+\rho_q)+7 \rho_q) \ln (\chi)  %
     - \beta_q (-1+\rho_q) \left(3 \rho ^2-4 \rho  \rho_q+\rho_q^2\right) 
    \ln \left(\frac{\chi +\chi_q}{1+\chi  \chi_q}\right) \Bigg\}\;.
\end{eqnarray}
\end{widetext}

\subsection{Light-Quark Scaling Functions $d_{k,\Pq}^{(1)}$ \label{app:quark-d}}
The light-quark scaling functions $d_{k,\Pq}^{(1)}$ at NLO accuracy for the Compton-like process
are given for the three projections $k=\{G,\,L,\,P\}$ by the following expressions:
\begin{widetext}
\begin{eqnarray}
d_{G,\Pq}^{(1)}(\eta,\xi) &=& \frac{1}{2592\pi\rho} \Bigg\{ 
     18 \left[ 8 \rho  \rho'-4 \rho'^2+\rho ^2 \left(-8+3 \rho'^2\right)\right]\; h_4(\chi,\chi') \nonumber\\
 &+& \beta  \left[ 5 \rho ^3+\rho ^2 (718-596 \rho')+32 \rho  \rho' (-26+5 \rho')+8 \rho'^2 (43+15 \rho')\right] 
     + \frac{9}{2} \rho  \left[16 \rho'+3 \rho  \left(-8+\rho ^2-4 \rho  \rho'\right)\right] \ln (\chi ) \nonumber\\
 &+& 12 \beta' \left[ \rho ^2 (38-23 \rho')+\rho  \rho' (-38+5 \rho')+\rho'^2 (16+5 \rho')\right] 
     \ln \left(\frac{\chi -\chi'}{1-\chi  \chi'}\right) \Bigg\} \;,\\
d_{L,\Pq}^{(1)}(\eta,\xi) &=& \frac{1}{216\pi\rho} \Bigg\{ 27 \rho'^3 (-\rho +\rho') h_4(\chi,\chi') 
     + \beta  \rho' \left[ 23 \rho ^2+2 (25-93 \rho') \rho'+19 \rho  (-2+7 \rho')\right] \nonumber\\
 &+& \frac{9}{2} \rho  \rho' \left(\rho ^2+3 \rho  \rho'-6 \rho'^2\right) \ln (\chi ) 
 + 6 \beta' (\rho -\rho') \rho' (-2+11 \rho') \ln \left(\frac{\chi -\chi'}{1-\chi  \chi'}\right) \Bigg\}\;, \\
d_{P,\Pq}^{(1)}(\eta,\xi) &=& \frac{1}{2592\pi\rho} \Bigg\{ 18 \left[ -4 \rho'^2+\rho  \rho' \left(8-3 \rho'^2\right)
      +\rho ^2 \left(-8+3 \rho'^2\right)\right]\;  h_4(\chi,\chi') \nonumber\\
 &+& \beta \left[ 5 \rho ^3+\rho ^2 (718-458 \rho')+8 (109-75 \rho') \rho'^2+20 \rho  \rho' (-53+41 \rho')\right] \nonumber\\
 &+& \frac{9}{2} \rho \left[ 16 \rho'+3 \rho  \left(-8+\rho ^2-2 \rho  \rho'+4 \rho'^2\right)\right] \ln (\chi ) \nonumber\\
 &-& 12 \beta' \left[ 2 \rho  (22-19 \rho') \rho'+\rho ^2 (-38+23 \rho')+\rho'^2 (-28+25 \rho')\right]
      \ln \left(\frac{\chi -\chi'}{1-\chi  \chi'}\right) \Bigg\}\;.
\end{eqnarray}
\end{widetext}

\section{Threshold Behavior of the Scaling Functions \label{app:threshold}}
In this Appendix, we list the subleading coefficients $a_{k,\tQED}^{(1,0)}$ 
and $a_{k,\tOK}^{(1,0)}$ appearing 
in the threshold limit of $c_{k,g}^{(1)}$ in Eq.~(\ref{eq:thr-approx}). 
We start, however, with briefly outlining of how to infer the threshold behavior
of the gluonic scaling function from our exact expressions for $c_{k,\Pg}^{(1)}(\eta,\xi)$.

The starting point is the phase space integration in Eq.~(\ref{eq:totalpartonic}) 
over the partonic variables $s_4$ and $t_1$, which we generically write as
\begin{equation}
\label{eq:thr-int0}
I = \int\limits_{t_{1,\min}}^{t_{1,\max}} dt_1 \int\limits_{\Delta}^{s_{4,\max}} ds_4\;f(t_1,s_4)\;.
\end{equation}
The limits of integration are explicitly given by
\begin{eqnarray}
t_{1,\min} &=& -\frac{s'} 2(1+\beta) = - \frac{s'\sqrt{\rho}} 2 \, \frac 1 {\sqrt\chi}\;, \nonumber\\
t_{1,\max} &=&-\frac{s'} 2(1-\beta) = - \frac{s'\sqrt{\rho}} 2 \, \sqrt\chi\;, \nonumber\\
s_{4,\max} &=& \frac{s}{s' t_1}\left(t_1+\frac{s'(1-\beta)}{2}\right)\left(t_1+\frac{s'(1+\beta)}{2}\right) \nonumber\\
 &=& s + \frac {s t_1}{s'} + \frac{s' m^2}{t_1}\;.
\end{eqnarray}

Next, one performs a series of straightforward but cumbersome transformations of variables and
integration limits and interchanges the order of integrations to finally arrive at 
\begin{equation}
\label{eq:thr-int}
I = -\int\limits_{a_{\min}}^{a_{\max}} da  \int\limits_{-1}^{1} db\,{\cal{J}}(a,b)\, f(t_1(a,b),s_4(a))
\end{equation}
where the Jacobian is given by
\begin{equation}
{\cal{J}}(a,b)=  -\frac{s'}{2s} \beta_4(a) \left(s(1-\sqrt\rho) - \Delta\right) \left[s-s_4(a)\right]\;.
\end{equation}
$s_4$ and $t_1$ are related to the new variables of integration by
\begin{eqnarray}
s_4(a) &=& a (s(1-\sqrt\rho) - \Delta) \;,\nonumber \\
t_1(a,b) &=& -\frac{s'}{2s}\left[s-s_4(a)\right]\left(1+b\beta_4(a)\right)
\end{eqnarray}
with
\begin{equation}
\beta_4(a) = \sqrt{1-\frac{4m^2 s}{\left[s-s_4(a)\right]^2}}
\end{equation}
and $\beta_4(0) = \beta$.
The limits of integration in (\ref{eq:thr-int}) are given by
\begin{eqnarray}
a_{\min} &=& \frac{\Delta}{s(1-\sqrt\rho) - \Delta} = \frac{\Delta}{s(1-\sqrt\rho)} + {\cal{O}}(\Delta^2)\;, \nonumber\\
a_{\max} &=& \frac{s(1-\sqrt\rho)}{s(1-\sqrt\rho) - \Delta} = 1 + {\cal{O}}(\Delta)\;.
\end{eqnarray}
Near threshold, one finds $a_{\min} = \frac{\Delta}{2m^2\beta^2} + {\cal{O}}(\beta^0)$.

Depending on the value of $s_4$, i.e., $s_4>\Delta$ or $s_4<\Delta$,
the kernel $f(t_1(a,b),s_4(a,b))$ in Eqs.~(\ref{eq:thr-int0}) and (\ref{eq:thr-int}) can be further decomposed into
contributions from hard and soft gluon radiation. This leads to
\begin{eqnarray}
\label{eq:thr-int-delta}
I &=& \int\limits_{t_{1,\min}}^{t_{1,\max}} \!\!\! dt_1 \!\!\!
      \int\limits_{0}^{s_{4,\max}} \!\!\! ds_4
      \left[ f_H(t_1,s_4)\Theta(s_4-\Delta) + f_S(t_1,\Delta)\delta(s_4)\right]\nonumber\\
 &=&  \int\limits_{0}^{1} da 
      \int\limits_{-1}^{1}db \;\frac{s'}{2}(1-\sqrt\rho)(s-s_4)\beta_4 \nonumber \\
 &\times& \left[f_H(t_1,s_4)\Theta\left(a-\frac{\Delta}{2m^2\beta^2}\right) 
                + f_S(t_1,\Delta)\frac{\delta(a)}{s(1-\sqrt\rho)}\right]\;.\nonumber\\
\end{eqnarray}
The threshold limit is obtained by expanding the kernel in $\beta$ before the integrations
in Eq.~(\ref{eq:thr-int-delta}) are performed. The results are then organized as in Eq.~(\ref{eq:thr-approx})
into LL, NLL, and subleading contributions, 
$a_{k,\tOK/\tQED}^{(1,2)}$, $a_{k,\tOK/\tQED}^{(1,1)}$, and $a_{k,\tOK/\tQED}^{(1,0)}$,
respectively.

To proceed, we define three auxiliary functions
\begin{eqnarray}
g_1(\chi_q) &=& \DiLog\left(-\frac{2 \chi_q}{1+\chi_q^2}\right) + \frac 1 2 \ln^2\left(\frac{2\chi_q}{1+\chi_q^2}\right)\nonumber\\
            &-& \frac {\pi^2} 2 - \frac 3 2 \ln^2(\chi_q)-2\beta_q \ln(\chi_q) \;,\\
g_2(\chi_q) &=& \ln(1+\chi_q)\left[\ln(1+\chi_q)-\ln(\chi_q)-\ln(2)\right] \nonumber \\
            &-& \ln(2)\left[2\ln(2)-\frac{1}{2} \ln(\chi_q)\right] \;,\\
g_3(\chi_q) &=& \ln(1+\chi_q) - \frac 1 2 \ln(\chi_q) - \frac 1 2 \ln(2)\;,
\end{eqnarray}
which allow us to write the OK and QED parts of the subleading coefficients $a_{k,\tOK}^{(1,0)}$ 
and $a_{k,\tQED}^{(1,0)}$ as follows:
\begin{widetext}
\begin{eqnarray}
\label{eq:at10ok}
a_{T,\tOK}^{(1,0)} = a_{P,\tOK}^{(1,0)} 
 &=& \frac{25}{8} + \frac 1 8 g_1(\chi_q) - g_2(\chi_q)
+ \frac{3-\rho_q}{2(2-\rho_q)}g_3(\chi_q) - \frac {15} 4 \ln(2) + \frac{5-2\rho_q}{8\beta_q}\ln(\chi_q) \nonumber \\
&&\hspace{7pt}+ \frac{\rho_q}{32(1-\rho_q)}\left[\pi^2+\ln^2(\chi_q)\right] \;,\\
\label{eq:al10ok}
a_{L,\tOK}^{(1,0)} &=& 
\frac {137}{36} + \frac{1+2\rho_q}{8(1-\rho_q)^2}g_1(\chi_q) - g_2(\chi_q) 
- \frac{1-(2-\rho_q)\rho_q}{(2-\rho_q)(1-\rho_q)}g_3(\chi_q)-\frac{19}{4}\ln(2) \nonumber \\
&&\hspace{7pt}+ \frac{(4-\rho_q)\rho_q}{24(1-\rho_q)^2}\left[\pi^2 + 3\ln(\chi_q)^2\right] \;,\\
\label{eq:at10qed}
a_{T,\tQED}^{(1,0)} = a_{P,\tQED}^{(1,0)} 
&=& \frac 1 {8(1-\rho_q)}\left[g_1(\chi_q)-\beta_q\ln(\chi_q)-\frac{2\pi^2}{3}\right]
+ \frac{(3-5\rho_q+2\rho_q^2)}{2(2-\rho_q)^2(1-\rho_q)}+\frac{4-\rho_q}{32(1-\rho_q)}\left[\pi^2+\ln(\chi_q)^2\right] \nonumber \\
&&\hspace{7pt}+\frac{9-5\rho_q}{8(2-\rho_q)}\;\\
\label{eq:al10qed}
a_{L,\tQED}^{(1,0)} &=& \frac{(1-6 \rho_q)}{8 (1-\rho_q)^2}\left[g_1(\chi_q)+\frac{\pi^2}{3}+\ln^2(\chi_q)\right] 
+ \frac{3+2\rho_q(5-(5-\rho_q)\rho_q)}{2(2-\rho_q)^2(1-\rho_q)}g_3(\chi_q) 
+ \frac{3+\rho_q}{\beta_q}\ln(\chi_q)\nonumber \\ 
&&\hspace{7pt}- \frac{3-2\rho_q}{8(2-\rho_q)} \;.
\end{eqnarray}
\end{widetext}
Finally, for completeness, 
in the limit of photoproduction, $q^2\to 0^-$, the expressions in Eqs.~(\ref{eq:at10ok}) and (\ref{eq:at10qed}) 
reduce to
\begin{eqnarray}
\lim_{q^2\to 0^-}a_{T,\tOK}^{(1,0)} &=& \frac 1 {48} \left[150 - 5 \pi^2 - 168\ln(2) +  96\ln^2(2)\right]
\;,\nonumber\\
\lim_{q^2\to 0^-}a_{T,\tQED}^{(1,0)} &=& \frac{1}{32} \left[-20+\pi ^2\right]\;.
\end{eqnarray}
Although Eqs.~(\ref{eq:al10ok}) and (\ref{eq:al10qed}) formally have
a non-vanishing limit $q^2\to 0^-$, there is, of course, no longitudinal partonic cross sections
for real photons, i.e., $\lim_{q^2\to 0^-} c_{L,g}=0$ to all orders, including
$c_{L,\Pg}^{(0),\text{thr}}$ in Eq.~(\ref{eq:thr-approx}).

\end{document}